\documentclass[twocolumn]{aastex631}
\usepackage{amsmath,graphicx,hyperref}
\usepackage{natbib,threeparttable,rotating}
\usepackage{ulem}
\usepackage{multirow}
\usepackage{longtable}
\newcommand{\etal}{et al.}

\newcommand{\hbeta}{H{$\beta$}}
\def\MgII{Mg\,{\sc ii}}
\def\CIV{C\,{\sc iv}}

\begin{document}

\title{The Sloan Digital Sky Survey Reverberation Mapping Project: The Black Hole Mass$-$Stellar Mass Relations at $0.2\lesssim z\lesssim 0.8$} 

\author[0000-0002-0311-2812]{Jennifer~I-Hsiu Li}
\affiliation{Department of Astronomy, University of Illinois at Urbana-Champaign, Urbana, IL 61801, USA}
\affiliation{Department of Astronomy, University of Michigan, Ann Arbor, MI, 48109, USA}

\author[0000-0003-1659-7035]{Yue Shen}
\affiliation{Department of Astronomy, University of Illinois at Urbana-Champaign, Urbana, IL 61801, USA}
\affiliation{National Center for Supercomputing Applications, University of Illinois at Urbana-Champaign, Urbana, IL 61801, USA}

\author[0000-0001-6947-5846]{Luis~C.~Ho}
\affiliation{Kavli Institute for Astronomy and Astrophysics, Peking University, Beijing 100871, People's Republic of China}
\affiliation{Department of Astronomy, School of Physics, Peking University, Beijing 100871, People's Republic of China}

\author{W.~N.~Brandt}
\affiliation{Department of Astronomy \& Astrophysics, The Pennsylvania State University, University Park, PA 16802, USA}
\affiliation{Institute for Gravitation and the Cosmos, The Pennsylvania State University, University Park, PA 16802, USA}
\affiliation{Department of Physics, The Pennsylvania State University, University Park, PA 16802, USA}

\author[0000-0001-9920-6057]{Catherine~J.~Grier} 
\affiliation{Department of Astronomy, University of Wisconsin-Madison, Madison, WI 53706, USA} 

\author[0000-0002-1763-5825]{Patrick B. Hall}
\affiliation{Department of Physics \& Astronomy, York University, 4700 Keele St., Toronto, ON M3J 1P3, Canada}

\author[0000-0002-0957-7151]{Y.~Homayouni}
\affiliation{Department of Astronomy \& Astrophysics, The Pennsylvania State University, University Park, PA 16802, USA}

\author[0000-0002-6610-2048]{Anton M. Koekemoer}
\affiliation{Space Telescope Science Institute, 3700 San Martin Dr., Baltimore, MD 21218, USA}

\author[0000-0001-7240-7449]{Donald P. Schneider}
\affiliation{Department of Astronomy \& Astrophysics, The Pennsylvania State University, University Park, PA 16802, USA}
\affiliation{Institute for Gravitation and the Cosmos, The Pennsylvania State University, University Park, PA 16802, USA}

\author[0000-0002-1410-0470]{Jonathan~R.~Trump}
\affiliation{Department of Physics, University of Connecticut, 2152 Hillside Rd Unit 3046, Storrs, CT 06269, USA}

\shorttitle{Black Hole Scaling Relations at $0.2\lesssim z\lesssim 0.8$}
\shortauthors{Li \etal}

\begin{abstract}
We measure the correlation between black-hole mass $M_{\rm BH}$ and host stellar mass $M_*$ for a sample of 38 broad-line quasars at $0.2\lesssim z\lesssim 0.8$ (median redshift $z_{\rm med}=0.5$). The black-hole masses are derived from a dedicated reverberation mapping program for distant quasars, and the stellar masses are estimated from two-band optical+IR HST imaging. {Most of these quasars are well centered within $\lesssim 1$\,kpc from the host galaxy centroid, with only a few cases in merging/disturbed systems showing larger spatial offsets.} Our sample spans two orders of magnitude in stellar mass ($\sim 10^9-10^{11}\,M_\odot$) and black-hole mass ($\sim 10^7-10^9\,M_\odot$), and reveals a significant correlation between the two quantities. {We find a best-fit intrinsic (i.e., selection effects corrected) $M_{\rm BH}-M_{\rm *,host}$ relation of $\log (M_{\rm BH}/M_{\rm \odot})=7.01_{-0.33}^{+0.23} + 1.74_{-0.64}^{+0.64}\log (M_{\rm *,host}/10^{10}M_{\rm \odot})$, with an intrinsic scatter of $0.47_{-0.17}^{+0.24}$~dex}. Decomposing our quasar hosts into bulges and disks, there is a similar $M_{\rm BH}-M_{\rm *,bulge}$ relation with slightly larger scatter, likely caused by systematic uncertainties in the bulge-disk decomposition. The $M_{\rm BH}-M_{\rm *,host}$ relation at $z_{\rm med}=0.5$ is similar to that in local quiescent galaxies, with negligible evolution over the redshift range probed by our sample. {With direct black-hole masses from reverberation mapping and a large dynamical range of the sample, selection biases do not appear to affect our conclusions significantly.} Our results, {along with other samples in the literature}, suggest that the locally-measured black-hole mass$-$host stellar mass relation is already in place at $z\sim 1$. 

\keywords{
black hole physics -- galaxies: active -- quasars: general -- surveys
}
\end{abstract}

\section{Introduction}
The observed scaling relations between supermassive black hole (BH) masses and the properties of their host galaxies (e.g., stellar mass and stellar velocity dispersion) {\it in the local Universe} are the foundation of modern BH$-$galaxy co-evolution models \citep[][and references therein]{Magorrian_etal_1998, Ferrarese_Merritt_2000, Gebhardt_etal_2000, Haring_Rix_2004, Gultekin_etal_2009, McConnell_Ma_2013, Kormendy_Ho_2013}. The tight correlations suggest that active galactic nuclei (AGNs) may play important roles in regulating star formation in the host galaxies via self-regulated BH growth and feedback processes \citep{Silk_Rees_1998, DiMatteo_etal_2005, Heckman_Best_2014}. Studying BH scaling relations beyond the local Universe is a key to understanding BH and galaxy (co-)evolution over cosmic history. 

Over the past two decades, various investigations have built an inventory of BH and host measurements to study the redshift evolution of BH$-$host relations up to $z\sim3$, including the BH mass$-$stellar velocity dispersion ($M_{\rm BH}-{\sigma}_{\rm *}$) relation \citep{Treu_etal_2004, Woo_etal_2006, Woo_etal_2010, Shen_etal_2015_Msigma, Park_etal_2015, Sexton_etal_2019}, the BH mass$-$bulge/host luminosity ($M_{\rm BH}-{L}_{\rm *, bulge/host}$) relation \citep{Peng_etal_2006a, Peng_etal_2006b, Decarli_etal_2010}, the BH mass$-$bulge/host stellar mass ($M_{\rm BH}-M_{\rm *, bulge/host}$) relation \citep{Jahnke_etal_2009, Bennert_etal_2011, Dong_etal_2016, Suh_etal_2020, Ding_etal_2021}, {as well as expanding the local baselines to include AGNs of different host properties and lower BH masses} \citep[e.g.,][]{Jiang_etal_2011a, Jiang_etal_2011b, Greene_etal_2008, Reines_Volonterni_2015, Bentz_etal_2018, Greene_etal_2020,  Bennert_etal_2021, Zhao_etal_2021}. Some groups found deviations from local scaling relations as a function of $z$ \citep[][]{Peng_etal_2006a, Peng_etal_2006b, Merloni_etal_2010, Woo_etal_2010, Park_etal_2015, Sexton_etal_2019} while others found similar BH$-$host relations as in the local Universe \citep[e.g.,][]{Jahnke_etal_2009, Suh_etal_2020, LiJunyao_etal_2021b, Ding_etal_2022, Silverman_etal_2022}, {which is also supported by the tight correlation between the BH accretion rate and star formation rate in bulge-dominated galaxies at $z=0.5-3$ \citep[e.g.,][]{Yang_etal_2019}.}

The measurements of BH mass$-$host relations can be challenging beyond $z\sim0.1$ for several reasons. First, direct BH mass measurements based on resolved stellar/gas dynamics are difficult to obtain beyond the local Universe where the BH sphere of influence cannot be readily resolved. Reverberation mapping \citep[RM;][]{Blandford_mcKee_1982, Peterson_2014} is the primary method of measuring BH masses for broad-line (BL) AGN beyond the local Universe, but RM is resource-intensive and only available for a small number of objects beyond $z\sim 0.1$ \citep[e.g.,][]{Bentz_etal_2013}. A secondary BH mass recipe, the single-epoch (SE) virial estimator, is based on the broad-line region (BLR) radius$-$luminosity relation (the $R-L$ relation) and can be easily adapted for large samples of BLAGNs at higher redshift. However, \citet{Shen_Kelly_2010} demonstrated that there is a statistical bias in SE BH masses for flux-limited samples from the uncertainties in these BH masses. In addition, the applicability of SE masses to the high-redshift and high-luminosity regime is not well understood, primarily because the local RM AGNs used to derive the $R-L$ relation is not representative of the general quasar population \citep[e.g.,][]{Shen_etal_2015_techoverview, FonsecaAlvarez_etal_2020}, and the extrapolated $R-L$ relations for broad \MgII\ and \CIV\ used for high-redshift BLAGNs are not as well-studied as the local $R-L$ relation based on broad \hbeta\ \citep[][]{Bentz_etal_2013}.

Host-galaxy properties are also difficult to measure as the unobscured AGN (where virial BH masses are feasible) usually far outshines the host galaxy. For imaging studies, high-resolution images, such as those from the Hubble Space Telescope (HST), are often necessary to robustly decompose the quasar and host light. {However, rigorous image analysis reveals that host galaxies of local AGNs ($z$\,$<$\,0.35) often consist of complex structures, including spiral arms, tidal and merger features, in addition to the main galaxy components (bars, bulges, and disks) \citep{Kim_etal_2017}. These complex structures are extremely challenging to measure even with HST at higher redshifts.}

Due to difficulties in obtaining BH mass and host properties, many studies are limited to specific samples that may introduce selection biases. Earlier studies were often restricted to the bright end of BLAGN, have small sample sizes and limited dynamical ranges in BH/host properties. \cite{Lauer_etal_2007} showed that over-massive BHs are favored in flux-limited studies due to the intrinsic scatter of the scaling relations. For a ``bottom-heavy'' galaxy luminosity function, there are more low-mass hosts than high-mass ones. However, more massive BHs are preferentially selected in a flux-limited sample based on AGN luminosity, resulting in an average offset in the BH mass$-$host relations, and a shallower slope than the true underlying relation. \citet{Schulze_Wisotzki_2011, Schulze_Wisotzki_2014} argued that additional selection biases could arise from the lack of knowledge in the relevant underlying distribution functions (e.g., the active fraction of AGNs, bulge properties) and their evolution with redshift. These biases can account for a large portion of, if not all, the redshift evolution reported in earlier investigations \citep{Schulze_Wisotzki_2011,Shen_etal_2015_Msigma}. 

\begin{figure}[t]
\centering
\includegraphics[width=0.45\textwidth]{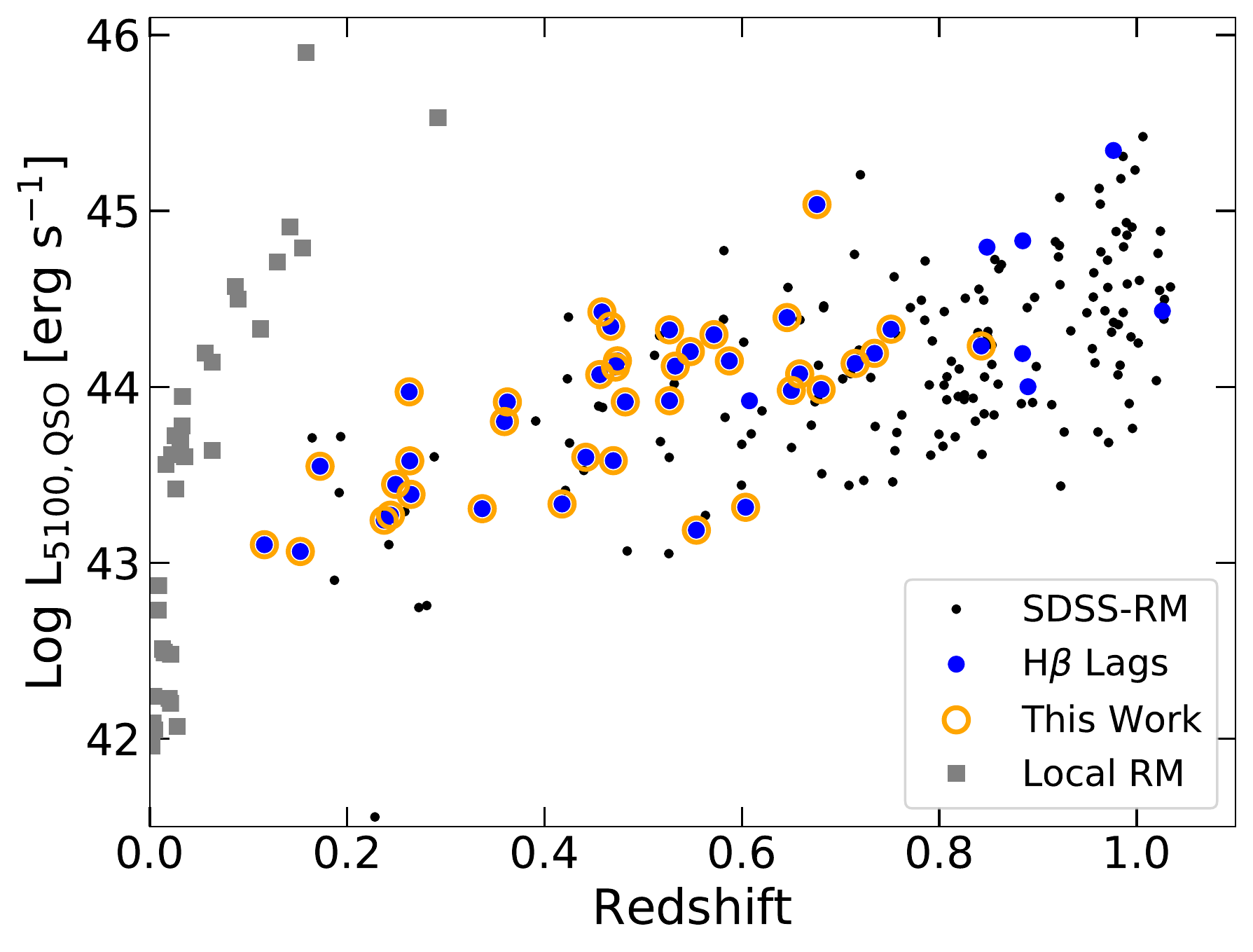}\\
\caption{Quasar luminosity and redshift distribution of our sample {(open orange circles)}, and a representative subset of the local RM sample (grey squares). The parent SDSS-RM sample (black dots) and those with \hbeta\,lags (blue dots) are also labeled for reference.}
\label{fig:sample}
\end{figure}

In this work, we study the BH scaling relations at $0.2\lesssim z\lesssim 0.8$ using the Sloan Digital Sky Survey Reverberation Mapping \cite[SDSS-RM,][]{Shen_etal_2015_techoverview} sample. The SDSS-RM sample has two major advantages in measuring the redshift evolution of BH$-$host galaxy relations: (1) the parent sample is a uniformly selected flux-limited BLAGN sample, thus the selection effects can be quantified and corrected, and (2) BH masses are available from direct RM \citep{Shen_etal_2016_firstlag,Grier_etal_2017,Grier_etal_2019,Homayouni_etal_2020}, rather than from SE, masses. We have acquired high-resolution imaging for the SDSS-RM sample with HST to measure the host galaxy color and luminosity in two bands, tracing young and old stellar populations, respectively. Our sample includes 38 sources \citep[10 included in a pilot study in][]{Li_etal_2021}, which is comparable in size to the local RM AGN sample used to calibrate the $R-L$ relation \citep{Bentz_etal_2013}, and has {sufficient statistics and dynamic range} in BH mass (and stellar mass) to characterize the redshift evolution of BH scaling relations over the redshift range of $0.2\lesssim z\lesssim 0.8$.

\begin{table*}[h]
\caption{Target Properties}
\label{tab:sample}
\begin{tabular}{cccccccc}
\hline\hline
RMID & R. A. (J2000) & Dec. (J2000) & $z$ & ${i}_{\rm psf}$ & ${{L}_{\rm 5100, QSO}}$ & log(${M}_{\rm BH, SE}$) & log(${{M}_{\rm BH, RM}}$) \\
 & (deg) & (deg) &  & (mag) & (erg\,${\rm s}^{-1}$) & ($\rm M_{\odot}$) & ($\rm M_{\odot}$) \\
\hline\hline
017 &213.3511 & 53.0908 & 0.4559 &19.21 & 43.9 & 8.36$\pm$0.04 & ${8.92}_{-0.19}^{+0.24}$ \\
033 &213.8848 & 52.8183 & 0.7147 &20.49 & 44.1 & 7.60$\pm$0.03 & ${7.23}_{-0.20}^{+0.23}$ \\
101 &213.0592 & 53.4296 & 0.4581 &18.84 & 44.4 & 7.89$\pm$0.004 & ${7.26}_{-0.19}^{+0.17}$ \\
160 &212.6719 & 53.3136 & 0.3593 &19.68 & 43.8 & 8.20$\pm$0.007 & ${7.85}_{-0.17}^{+0.18}$ \\
177 &214.3525 & 52.5069 & 0.4818 &19.56 & 44.0 & 8.43$\pm$0.03 & ${7.57}_{-0.20}^{+0.55}$ \\
191 &214.1899 & 53.7463 & 0.4418 &20.45 & 43.6 & 7.55$\pm$0.01 & ${6.90}_{-0.16}^{+0.22}$ \\
229 &212.5752 & 53.4937 & 0.4696 &20.27 & 43.6 & 8.00$\pm$0.07 & ${7.65}_{-0.20}^{+0.17}$ \\
265 &215.0995 & 53.2681 & 0.7343 &20.65 & 44.2 & 8.31$\pm$0.02 & ${8.58}_{-0.26}^{+0.23}$ \\
267 &212.8030 & 53.7520 & 0.5872 &19.62 & 44.1 & 7.92$\pm$0.02 & ${7.41}_{-0.17}^{+0.17}$ \\
272 &214.1071 & 53.9107 & 0.2628 &18.82 & 43.9 & 7.82$\pm$0.02 & ${7.58}_{-0.21}^{+0.18}$ \\
300 &214.9213 & 53.6138 & 0.6457 &19.49 & 44.5 & 8.19$\pm$0.02 & ${7.60}_{-0.20}^{+0.17}$ \\
301 &215.0427 & 52.6749 & 0.5477 &19.76 & 44.1 & 8.53$\pm$0.09 & ${8.64}_{-0.22}^{+0.25}$ \\
305 &212.5178 & 52.5281 & 0.5266 &19.50 & 44.2 & 7.92$\pm$0.01 & ${8.32}_{-0.16}^{+0.16}$ \\
316 &215.2185 & 52.9396 & 0.6760 &18.03 & 45.0 & 8.50$\pm$0.006 & ${}^{*}{7.55}_{-0.17}^{+0.17}$ \\
320 &215.1605 & 53.4046 & 0.2647 &19.47 & 43.4 & 8.06$\pm$0.02 & ${7.67}_{-0.18}^{+0.18}$ \\
338 &214.9818 & 53.6687 & 0.4177 &20.08 & 43.4 & 8.36$\pm$0.05 & ${7.69}_{-0.24}^{+0.27}$ \\
371 &212.8476 & 52.2255 & 0.4719 &19.57 & 44.1 & 8.13$\pm$0.02 & ${7.38}_{-0.16}^{+0.16}$ \\
377 &215.1814 & 52.6032 & 0.3368 &19.77 & 43.4 & 7.90$\pm$0.03 & ${7.20}_{-0.16}^{+0.16}$ \\
392 &215.3012 & 52.6965 & 0.8425 &20.44 & 44.3 & 8.19$\pm$0.04 & ${8.22}_{-0.18}^{+0.19}$ \\
457 &213.5714 & 51.9563 & 0.6037 &20.29 & 43.4 & 8.10$\pm$0.1 & ${8.03}_{-0.21}^{+0.18}$ \\
519 &214.3012 & 51.9460 & 0.5538 &21.54 & 43.2 & 7.36$\pm$0.08 & ${}^{*}{7.38}_{-0.19}^{+0.18}$ \\
551 &212.9461 & 51.9388 & 0.6802 &21.52 & 44.0 & 7.66$\pm$0.03 & ${6.95}_{-0.19}^{+0.19}$ \\
589 &215.2053 & 52.1815 & 0.7510 &20.74 & 44.4 & 8.52$\pm$0.02 & ${9.00}_{-0.18}^{+0.18}$ \\
601 &212.2685 & 54.0623 & 0.6585 &20.10 & 44.1 & 9.06$\pm$0.05 & ${8.45}_{-0.24}^{+0.36}$ \\
622 &212.8133 & 51.8692 & 0.5716 &19.55 & 44.3 & 8.22$\pm$0.08 & ${7.94}_{-0.16}^{+0.19}$ \\
634 &212.8995 & 51.8346 & 0.6500 &20.76 & 44.0 & 7.46$\pm$0.03 & ${7.56}_{-0.24}^{+0.27}$ \\
645 &215.1658 & 52.0666 & 0.4738 &19.78 & 44.1 & 8.22$\pm$0.01 & ${7.57}_{-0.18}^{+0.16}$ \\
694 &214.2778 & 51.7278 & 0.5324 &19.62 & 44.2 & 7.59$\pm$0.008 & ${6.70}_{-0.17}^{+0.35}$ \\
720 &211.3251 & 53.2583 & 0.4670 &19.03 & 44.3 & 8.14$\pm$0.007 & ${7.74}_{-0.18}^{+0.22}$ \\
767 &214.2122 & 53.8658 & 0.5266 &20.23 & 43.9 & 7.51$\pm$0.04 & ${}^{*}{7.63}_{-0.16}^{+0.17}$ \\
772 &215.3996 & 52.5275 & 0.2491 &18.87 & 43.4 & 7.63$\pm$0.02 & ${6.60}_{-0.22}^{+0.22}$ \\
775 &211.9961 & 53.7999 & 0.1725 &17.91 & 43.5 & 7.93$\pm$0.008 & ${7.67}_{-0.24}^{+0.39}$ \\
776 &212.0504 & 53.8842 & 0.1161 &17.98 & 43.1 & 7.80$\pm$0.007 & ${7.26}_{-0.19}^{+0.17}$ \\
779 &214.8474 & 54.3671 & 0.1525 &19.10 & 43.1 & 7.43$\pm$0.01 & ${7.18}_{-0.17}^{+0.17}$ \\
781 &215.2647 & 51.9721 & 0.2634 &19.31 & 43.6 & 7.77$\pm$0.01 & ${7.89}_{-0.16}^{+0.16}$ \\
782 &213.3290 & 54.5340 & 0.3623 &18.89 & 43.9 & 8.01$\pm$0.009 & ${7.51}_{-0.18}^{+0.16}$ \\
790 &214.3720 & 53.3074 & 0.2374 &18.67 & 43.3 & 8.43$\pm$0.01 & ${8.28}_{-0.23}^{+0.48}$ \\
840 &214.1881 & 54.4280 & 0.2439 &18.63 & 43.2 & 8.29$\pm$0.03 & ${7.93}_{-0.20}^{+0.21}$ \\
\hline\hline
 \end{tabular}
 \tablecomments{RM black hole masses are based on \hbeta\, lags from \cite{Grier_etal_2017}, except for RM767, which is based on the \MgII\ lag from \citet{Shen_etal_2016_firstlag}. $L_{\rm 5100,QSO}$ are from \citet{Shen_etal_2015_Msigma}, and are the host light-subtracted quasar continuum luminosity at restframe 5100\,\AA. {The single-epoch BH mass uncertainties are $1\sigma$ measurement errors only, but SE BE masses are typically dominated by systematic uncertainty of $\sim$0.5\,dex. The RM BH mass uncertainties also include 0.16\,dex systematic uncertainty following \citet{Grier_etal_2017}.}  $M_{\rm BH}$ for RM316, RM519, and RM767 (labeled with asterisks) are updated from \citet{Grier_etal_2017} and \citet{Li_etal_2021} as described in Section \ref{sec:mbh}.  }
\end{table*}

This paper is organized as follows. We describe our data and analysis in Section \ref{sec:obs}. The main results are presented in Section \ref{sec:results}. We discuss our results in Section \ref{sec:discussion} and conclude in Section \ref{sec:conclusions}. Throughout this paper we adopt a flat $\Lambda$CDM cosmology with $\Omega_M=0.3$ and $H_0=70\,{\rm km\,s^{-1}\,Mpc^{-1}}$. All host-galaxy measurements refer to the stellar population only.

\section{Observation and Data Analysis}\label{sec:obs}
\subsection{Sample}
Our sample consists of 38 SDSS-RM quasars at $0.2\lesssim z\lesssim0.8$ (median redshift $z_{\rm med}=0.5$) with RM-based BH masses; 37 of these RM masses were based on the broad \hbeta\ line \citep{Shen_etal_2016_firstlag, Grier_etal_2017}, with one source (RM767) based on the broad \MgII\ line \citep{Shen_etal_2016_firstlag}. Ten sources in our HST sample were studied in a pilot program \citep[][]{Li_etal_2021}; 28 sources are presented in this work for the first time. Among the 44 quasars with \hbeta\ RM BH masses in \citet{Grier_etal_2017}, seven sources beyond $z\sim 0.8$ were excluded from the HST programs to ensure more robust host-galaxy measurements and to avoid unknown selection biases, as the lag-detection fraction at $z\gtrsim0.8$ is significantly lower than that at lower redshifts {(e.g., see Figure \ref{fig:sample})}. Figure \ref{fig:sample} presents the redshift and luminosity distribution of our sample, and Table \ref{tab:sample} summarizes the physical properties of these objects.

\subsection{Black-Hole Masses} \label{sec:mbh}
Reverberation mapping determines BH masses by measuring the time delay in variability between the continuum and broad emission lines. The time delay corresponds to the light travel time between the continuum-emitting accretion disk and the BLR. Assuming the BLR is virialized, a BH mass can be calculated using the average time lag ($\tau$) and the width of the broad emission line ($\Delta V$) via the equation: 
\begin{equation}
M_{\rm BH} = f \frac{c {\tau} {\Delta V}^{2}}{G}, 
\end{equation}
where $G$ is the gravitational constant and $f$ is a dimensionless factor of unity order that accounts for BLR geometry, kinematics, and inclination. The line width $\Delta V$ can be estimated from either the full-width half-maximum (FWHM) or the line dispersion ($\sigma_{\rm line}$) of the broad line measured from the mean or RMS spectra \citep[e.g.,][]{Wang_etal_2019}. 

For the majority of our sources, we adopt the RM black hole masses from \cite{Grier_etal_2017} computed using a consant virial coefficient of $f=$4.47 based on $\sigma_{\rm line}$ measured from the RMS spectra (equivalent to $f=1.12$ when using the FWHM for $\Delta V$). For two of our sources, RM316 and RM519, the original $\sigma_{\rm line}$ measurements in \citet{Grier_etal_2017} based on the first-year SDSS-RM spectroscopy are significantly overestimated; we adopt updated $\sigma_{\rm line}$ measurements based on the 4-year SDSS-RM spectroscopy for these two objects. For RM767, \cite{Shen_etal_2016_firstlag} identified a lag between the continuum and broad \MgII\, line during the first-year monitoring. However, the lag significance is reduced in the more recent analysis in \cite{Homayouni_etal_2020} using 4-year light curves, as the broad \MgII\ line does not display strong response to the continuum in the following years. We adopt the \cite{Shen_etal_2016_firstlag} \MgII\ lag for RM767, and use its $\sigma_{\rm line}$ measured from the RMS spectrum to derive a BH mass. {The BH mass uncertainties are calculated by propagating the statistical uncertainties of the lag and line width measurements, and then adding a systematic uncertainty of 0.16\,dex, which is the scatter estimated from repeated RM measurements in local RM campaigns \citep{Fausnaugh_etal_2017}. However, the adopted BH mass uncertainty is still an underestimation, as it does not account for the intrinsic scatter in the virial coefficient for individual systems, which could lead to additional BH mass uncertainties of as large as $\sim$\,0.3\,dex. The BH masses are tabulated in Table \ref{tab:sample} (with updates from earlier work indicated by an asterisk). }

\subsection{HST Imaging Analysis}

The HST observations for the 28 new objects were conducted between 2019 December 23 and 2021 June 09 in Cycle 27 (GO-15849, PI: Shen). Our observational design is identical to the pilot program (GO-14109, PI: Shen): each target was observed with two dedicated orbits, one in UVIS filters (F606W for $z<0.6$ and F814W for $z>0.6$) and one in IR filters (F110W for $z<0.6$ and F140W for $z>0.6$), which are chosen to cover similar rest-frame wavelengths at different redshifts. Two additional orbits were used to observe the white dwarf EGGR-26 to construct the point spread function (PSF) models in all bands used for this program. All observations were performed in dithered patterns (three-point dithering for UVIS filters and four-point dithering for IR filters) to improve PSF sampling. The data were processed using standard HST calibration procedures and geometrically corrected and dither-combined with {\it astrodrizzle}. The final image sampling is 0$''$.033\,pixel$^{-1}$ for the UVIS F606W/F814W images and 0$''$.066\,pixel$^{-1}$ for the IR F110W/F140W images, which correspond to $\sim0.2$ and $\sim0.4$\,kpc at $z=0.5$. The FWHM of the PSF is $\sim2.2$\,pixels for the IR images and $\sim1.8$\,pixels for the UVIS images.

\begin{figure*}[t]
\includegraphics[width=0.5\textwidth]{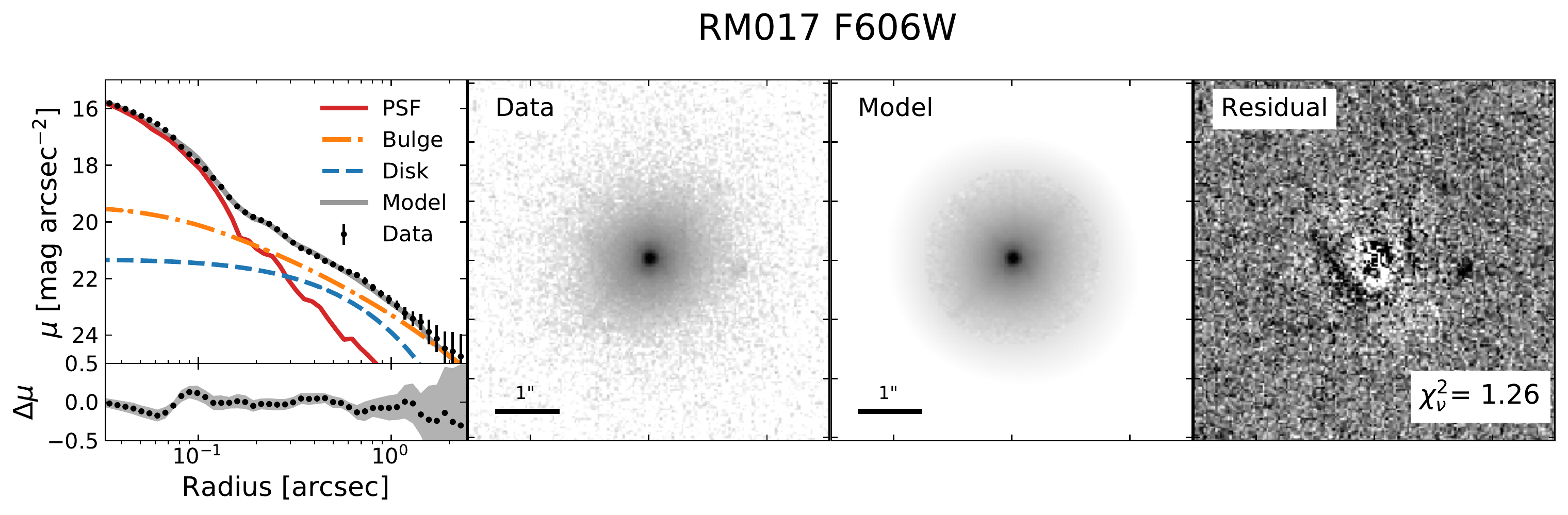}
\includegraphics[width=0.5\textwidth]{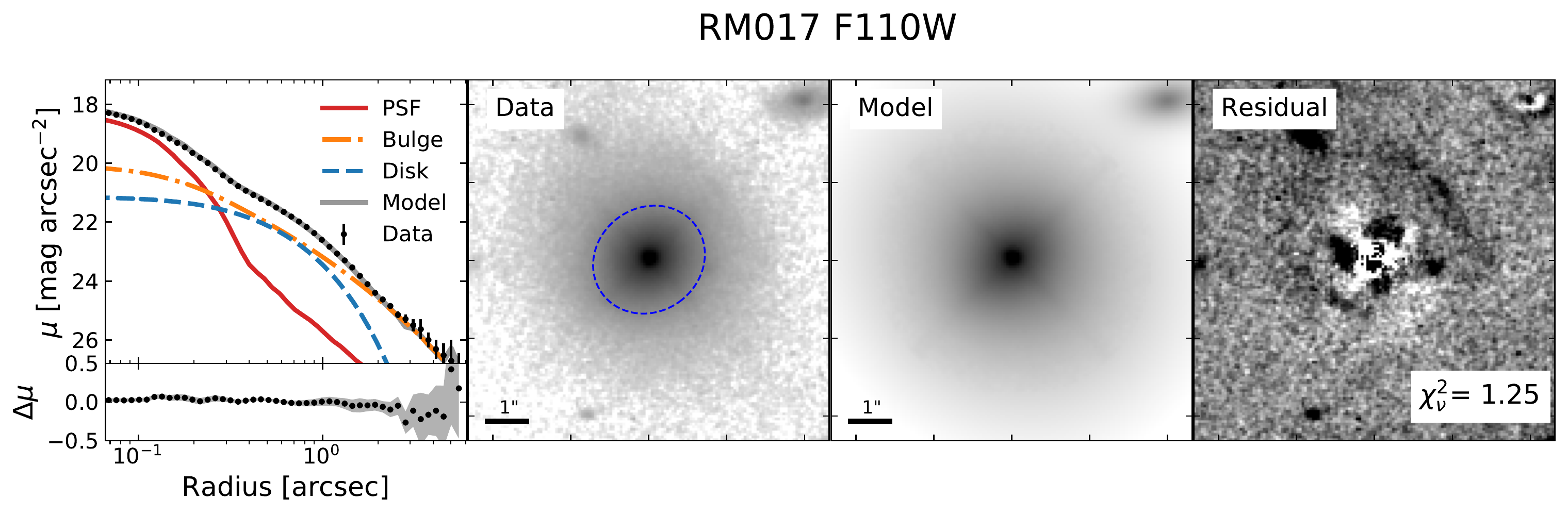}\\
\includegraphics[width=0.5\textwidth]{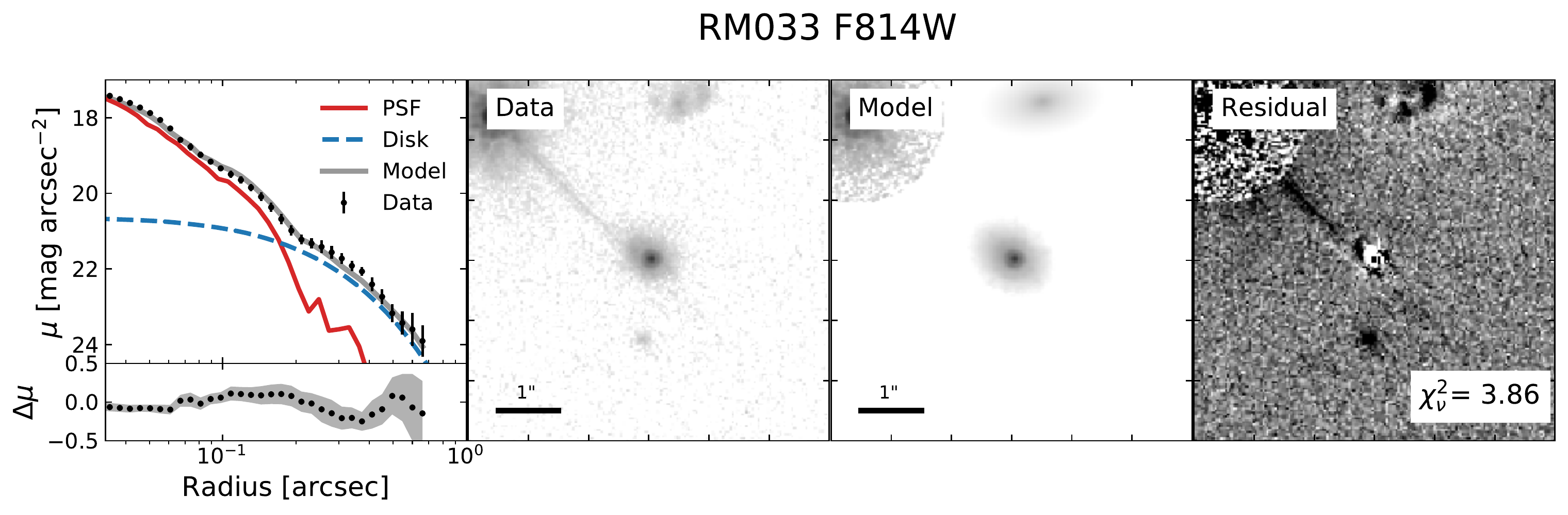}
\includegraphics[width=0.5\textwidth]{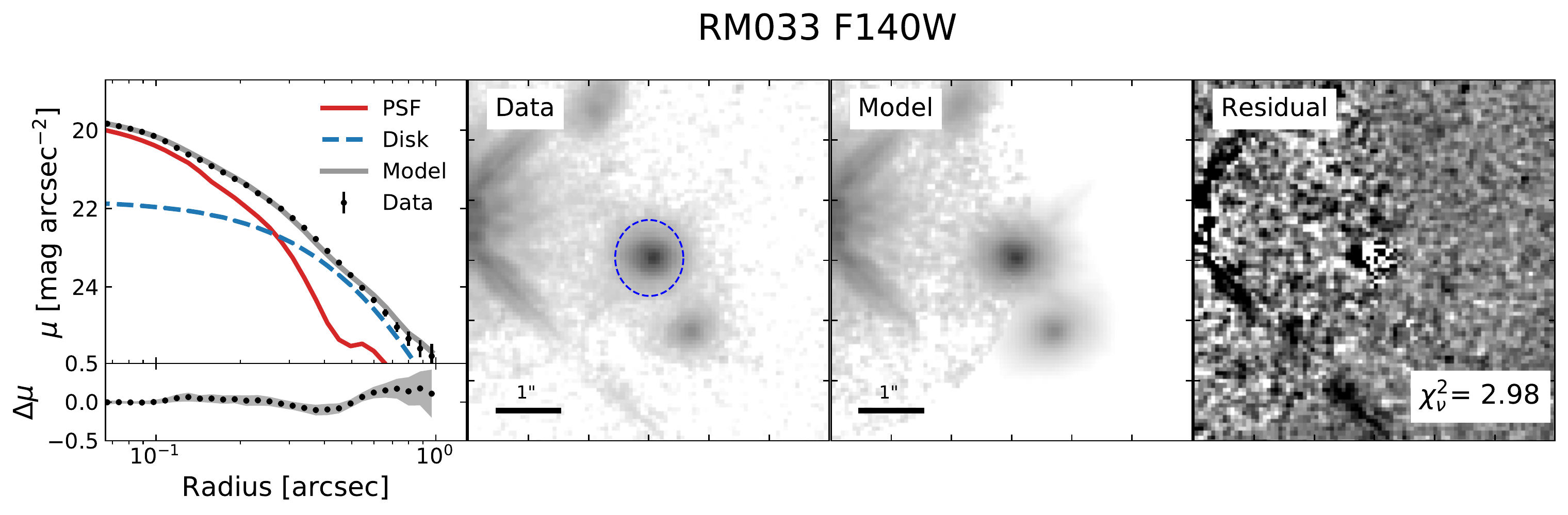}\\
\includegraphics[width=0.5\textwidth]{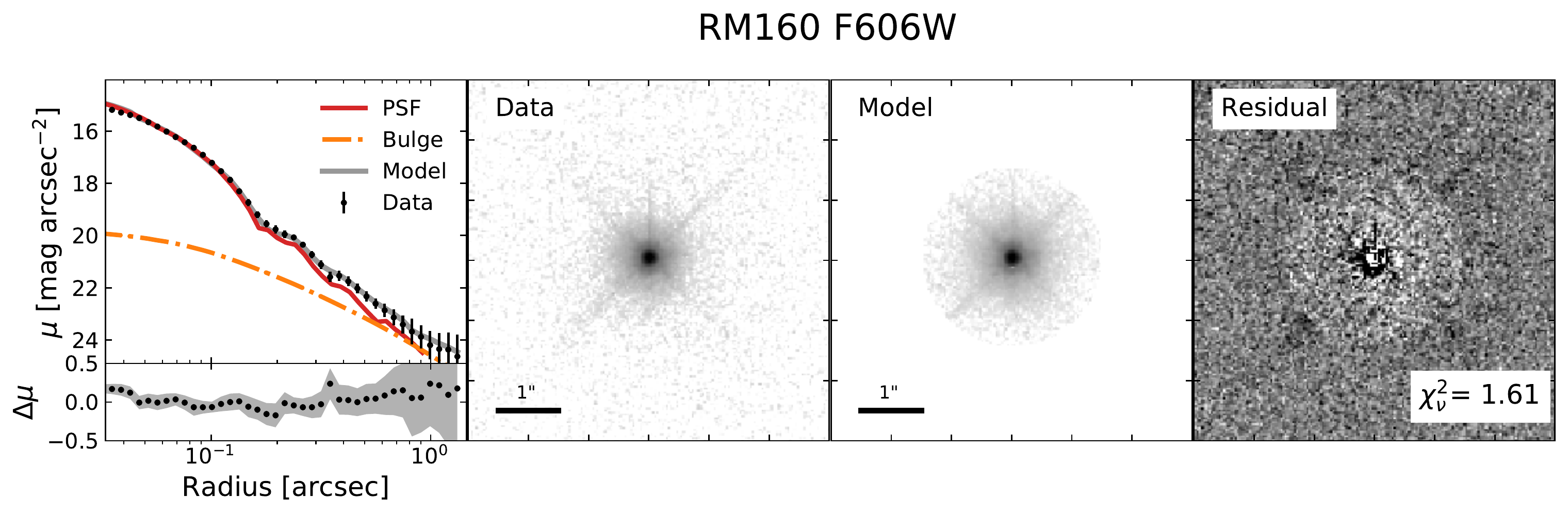}
\includegraphics[width=0.5\textwidth]{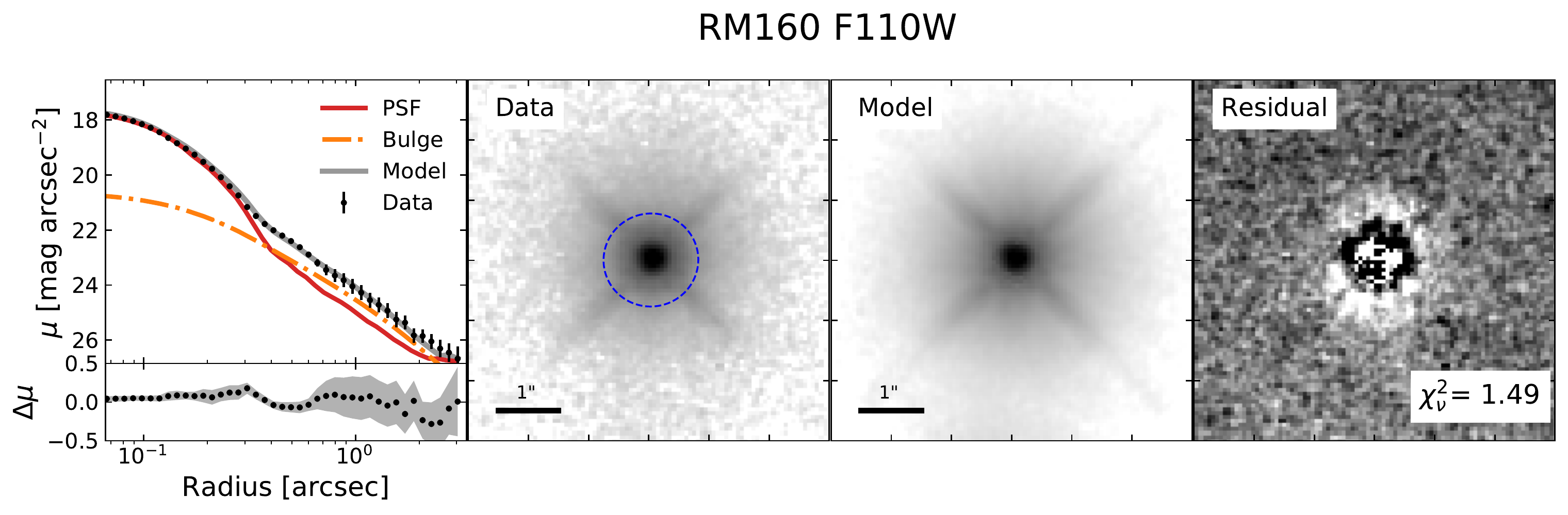}\\
\caption{Examples of surface-brightness decomposition of three quasars with {\tt GALFIT}, from top to bottom are sources that are best-fitted by a PSF+bulge+disk, PSF+disk, and PSF+bulge model, respectively. The left panels are the surface-brightness profiles of the data (black dots), the model (grey solid line) and each modeled component (red solid lines for PSFs, orange dotted-dash lines for bulges ($n$=4), blue dash lines for exponential disks ($n$=1), and purple dotted lines for truncated rings in RM775 and RM790). The radial profiles are directly-measured from the {\tt GALFIT} decomposed models and the HST images with isophote fitting. The leftmost, bottom sub-panel for each object is the residual of the surface-brightness profile, {with the rms along the isophote elliptical plotted in grey}. The three images on the right are (from left to right) the HST image, the {\tt GALFIT} model, and the residual. The blue ellipse in the HST image (IR only) encloses the area above 3$\sigma$ sky background in the best-fitting model. The residual images display the 1$^{\rm st}$ to 99$^{\rm th}$ percentiles (with linear stretch) of the residual values to provide better visual contrast. The reduced ${\chi}^{2}$ of the model is labeled in the lower right corner of each residual image. The full figure set is available online. }
\label{fig:galfit_example}
\end{figure*}

For RM177, our HST program only covers the IR band, because this object was observed in UVIS (F606W and F814W) from a previous HST program \citep[GO-10134, PI: Davis, ][]{Davis_etal_2007}. We processed the individual UVIS exposures from this earlier program following the same procedures for our HST program, and the final imaging sampling is 0$''$.05\,pixel$^{-1}$ ($\sim0.3$\,kpc at $z=0.5$). We use a field star in the same field of view as the PSF model for the UVIS images of RM177. 

We then follow the procedures in \cite{Li_etal_2021} and perform 2D image decomposition to separate the quasar and host light using {\tt GALFIT} \citep{galfit}, but with two modifications. First, we allow each system to be fitted by a PSF+disk model (S\'ersic index $n=1$), in addition to the PSF+bulge model ($n=4$) and PSF+bulge+disk model. Upon our analysis with the full sample, we identified several sources whose hosts best fitted by an exponential disk, rather than a bulge or bulge+disk model as used in \citet{Li_etal_2021}. Second, we revise our model-selection criteria using reduced-${\chi}^{2}$ calculated from a small region surrounding the target. By default, {\tt GALFIT}'s reduced-${\chi}^{2}$ is calculated from the entire image analysis area (e.g., roughly 10$''\times$10$''$) where $>60\%$ of all the pixels are background, so the reduced-${\chi}^{2}$ can change based on the chosen image size, and is largely determined by the accuracy of the background estimation. To assess the quality of the fit, we calculate $r{\chi}^{2}_{e}$, the reduced-$\chi^{2}$ within the best-fit ellipse at 3$\sigma$ sky background {(estimated by {\tt GALFIT})} around the source. If $\Delta(r{\chi}^{2}_{e})>5$ {(threshold chosen by visual inspection of the data)} between the PSF+bulge+disk model and the 2-component models (PSF+bulge or PSF+disk), we consider there is strong evidence that the additional component is necessary and adopt the three-component model; otherwise, we select the two-component model (PSF+bulge or PSF+disk) with the smaller $r{\chi}^{2}_{e}$ as the best-fit model.

To briefly summarize our fitting procedure, we first fit the IR images with three different models: PSF+bulge (S\'ersic index $n=4$), PSF+disk ($n=1$), and PSF+bulge+disk. The fit is considered successful when the best-fit parameters are within reasonable ranges (i.e., the effective radius of the S\'ersic component $R_{e}>1$ pixel, axis ratio $q>0.01$), which is to prevent introducing additional components fitting for mismatched PSF or other small-scale features. While the galaxy may not be a perfect bulge or disk, \cite{Kim_etal_2008a} showed that fixing the S\'ersic index results in more accurate flux recovery during host decomposition when the host galaxy is faint. We use $\Delta(r{\chi}^{2}_{e})$  to select the best-fit model from the successful PSF+bulge, PSF+disk, and PSF+bulge+disk models. {In addition to the quasar+host, we fit additional PSF and/or S\'ersic models for nearby objects to ensure the host decomposition and sky background estimation are not strongly affected by nearby objects (e.g., see RM033, RM101, RM694, RM776, etc, for examples).}

We visually inspect all the {\tt GALFIT} images and manually adjust the {\tt GALFIT} models only when necessary. Upon visual inspection, the background in RM776 is high due to a nearby bright object, and adding another component improves the fitting of its surface-brightness profile significantly, so we adopt a three-component model for RM776. {RM775 and RM790 display extended truncated ring features in the residual images of the PSF+bulge+disk model, so a fourth component (an inner-truncated disk) was added to ensure robust flux recovery for the host. The truncated disk in RM775 is also fitted with Fourier modes to account for the irregular ellipsoid shape. However, we only include the main disk component in the PSF+bulge+disk model, and not the truncated disk, for estimating the final photometry for the disks.} Finally, we fit the flux of each component in the UVIS images by fixing the shape and structural parameters (S\'ersic index, effective radius, ellipticity, and position angle) to the best-fit model in the IR images. For the sources that preferred the three-component model in the IR image, we check if the three components in the UVIS image converge on similar relative positions as in the IR image, which ensures the model is fitting the same physical structures in the two bands. The bulge and disk components of two sources, RM267 and RM316, failed to converge at similar central positions, so the two-component model (PSF+bulge) is adopted instead. Fig.~\ref{fig:galfit_example} presents a few examples of our {\tt GALFIT} decomposition, and the {\tt GALFIT} decomposition results are tabulated in Table \ref{tab:galfit}. The complete figure set, data, PSF templates, and {\tt GALFIT} decomposition models are available via \url{ftp://quasar.astro.illinois.edu/public/sdssrm/paper_data/Li_2023_HST_host}.

During our analysis of the full HST sample, we discovered an error in our {\tt GALFIT} analysis in the pilot study \citep{Li_etal_2021}. The {\tt ncombine} parameter was input incorrectly, {which caused the sigma image produced by {\tt GALFIT} to be overestimated by a factor of $\sim$\,4 in areas dominated by emission} (see {\tt GALFIT} user manual, Equation 33). The error mainly affects the estimation of $\chi^2$, but does not change the fitting results, i.e., all fitted parameters are consistent with the results with the correct sigma images within the uncertainties. We include updated measurements for the 10 objects in the pilot study in Table \ref{tab:galfit}.

{\tt GALFIT} only accounts for statistical uncertainties between the data and the model, and does not take into account PSF mismatches or complex spatial structures. There are three major sources of flux uncertainties: (1) the {temporal variability of the HST PSF (derived from the difference between the dedicated PSF observation and field stars in science observations}, $\sim0.07$\,mag in UVIS and $\sim0.03$\,mag in IR), (2) the deviation between the {\tt GALFIT} model and the image ($\sim0.02$\,mag in UVIS and $\sim0.005$\,mag in IR), and (3) fixing the S\'ersic index ($\sim0.05$\,mag for PSF and $\sim0.2$\,mag for the host/bulge/disk). We combine these flux uncertainties and adopt typical values of 0.1 and 0.25\,mag as the final uncertainties for the PSF and galaxy (bulge, disk, or galaxy) flux measurements in all bands, respectively. These final uncertainties are consistent with those in our pilot study and similar observations and simulations in the literature \citep[e.g.,][]{Kim_etal_2008a, Jahnke_etal_2009, Park_etal_2015, Bentz_etal_2018}. See \cite{Li_etal_2021} for additional technical details on the flux uncertainty budget.

\begin{table*}
\caption{{Galaxy Decomposition Results}}
\setlength\tabcolsep{6pt}
\label{tab:galfit}
\begin{tabular}{c|crrrrrrrr}
\hline\hline
RMID & Comp. & ${\rm {mag}_{UVIS}}$ & ${\rm {mag}_{IR}}$ & r ($''$) & n & q & P. A. & ${\rm {r\chi^{2}}_{UVIS}}$ & ${\rm {r\chi^{2}}_{IR}}$\\
\hline\hline									
017& PSF & 20.15 & 21.19 & & & & & 1.26 & 1.25\\
 &Bulge & 19.98 & 19.88 & 1.50 &4 & 0.89 & 149.0 & & \\
 &Disk & 21.58 & 21.11 & 0.63 &1 & 0.67 & 53.2 & & \\
\hline
033& PSF & 21.69 & 22.59 & & & & & 3.86 & 2.98\\
 &Disk & 22.36 & 23.01 & 0.27 &1 & 0.69 & -60.0 & & \\
\hline
101& PSF & 19.41 & 20.53 & & & & & 1.43 & 1.17\\
 &Bulge & 21.17 & 21.14 & 0.71 &4 & 0.88 & -158.4 & & \\
\hline
160& PSF & 19.35 & 20.52 & & & & & 1.61 & 1.49\\
 &Bulge & 21.28 & 21.27 & 0.67 &4 & 0.99 & 79.0 & & \\
\hline
177& PSF & 20.06 & 21.60 & & & & & 1.76 & 1.37\\
 &Disk & 21.53 & 21.53 & 0.53 &1 & 0.50 & 151.3 & & \\
\hline
191& PSF & 23.02 & 24.29 & & & & & 1.28 & 2.12\\
 &Bulge & 23.22 & 22.26 & 0.50 &4 & 0.17 & -130.1 & & \\
 &Disk & 20.81 & 20.91 & 1.33 &1 & 0.54 & -154.5 & & \\
\hline
229& PSF & 21.49 & 22.49 & & & & & 1.24 & 1.89\\
 &Bulge & 24.70 & 23.22 & 0.31 &4 & 0.21 & -118.5 & & \\
 &Disk & 21.66 & 21.61 & 0.78 &1 & 0.69 & -134.4 & & \\
\hline
265& PSF & 21.57 & 22.55 & & & & & 1.24 & 1.36\\
 &Disk & 20.84 & 21.01 & 1.36 &1 & 0.95 & 118.9 & & \\
\hline
267& PSF & 20.25 & 21.64 & & & & & 1.36 & 1.70\\
 &Bulge & 21.11 & 21.12 & 0.37 &4 & 0.71 & 67.1 & & \\
\hline
272& PSF & 19.10 & 20.31 & & & & & 1.26 & 1.62\\
 &Bulge & 21.47 & 21.47 & 0.31 &4 & 0.56 & 144.0 & & \\
 &Disk & 21.01 & 21.43 & 0.82 &1 & 0.34 & 123.0 & & \\
\hline
300& PSF & 20.39 & 21.56 & & & & & 1.36 & 2.32\\
 &Disk & 21.75 & 22.33 & 0.40 &1 & 0.97 & -123.1 & & \\
\hline
301& PSF & 21.02 & 22.42 & & & & & 1.32 & 1.64\\
 &Bulge & 22.58 & 21.66 & 0.32 &4 & 0.34 & 37.0 & & \\
 &Disk & 21.37 & 21.38 & 1.44 &1 & 0.69 & 47.4 & & \\
\hline
305& PSF & 20.39 & 21.31 & & & & & 1.48 & 1.39\\
 &Bulge & 21.17 & 20.90 & 0.68 &4 & 0.88 & 71.6 & & \\
\hline
316& PSF & 19.07 & 20.35 & & & & & 1.41 & 2.38\\
 &Bulge & 20.88 & 21.10 & 0.96 &4 & 0.68 & 27.4 & & \\
\hline
320& PSF & 20.75 & 21.81 & & & & & 1.27 & 2.73\\
 &Bulge & 21.80 & 21.65 & 0.21 &4 & 0.66 & -138.5 & & \\
 &Disk & 19.88 & 20.02 & 2.41 &1 & 0.34 & 154.5 & & \\
\hline
338& PSF & 21.14 & 21.96 & & & & & 1.39 & 1.47\\
 &Disk & 20.94 & 21.18 & 0.68 &1 & 0.89 & -50.5 & & \\
\hline
371& PSF & 20.29 & 21.21 & & & & & 1.35 & 2.09\\
 &Bulge & 20.88 & 20.77 & 1.11 &4 & 0.86 & 45.0 & & \\
\hline
377& PSF & 22.64 & 23.32 & & & & & 1.15 & 1.59\\
 &Bulge & 21.06 & 20.91 & 0.33 &4 & 0.52 & 100.8 & & \\
 &Disk & 21.36 & 21.52 & 1.12 &1 & 0.92 & -174.1 & & \\
\hline
392& PSF & 22.62 & 23.18 & & & & & 1.29 & 1.42\\
 &Bulge & 21.94 & 22.10 & 0.60 &4 & 0.71 & 151.6 & & \\
\hline\hline
 \end{tabular}
\end{table*}
 
\begin{table*}[t]
\movetableright=-0.4in
\setcounter{table}{1}
\caption{(Continued).}
\setlength\tabcolsep{6pt}
\label{tab:galfit2}
\begin{tabular}{c|crrrrrrrr}
\hline\hline
RMID & Comp. & ${\rm {mag}_{UVIS}}$ & ${\rm {mag}_{IR}}$ & r ($''$) & n & q & P. A. & ${\rm {r\chi^{2}}_{UVIS}}$ & ${\rm {r\chi^{2}}_{IR}}$\\
\hline		
\hline
457& PSF & 22.71 & 23.56 & & & & & 1.20 & 1.31\\
 &Bulge & 22.43 & 22.21 & 0.78 &4 & 0.75 & 139.4 & & \\
\hline
519& PSF & 22.76 & 23.59 & & & & & 1.19 & 1.37\\
 &Bulge & 23.55 & 23.51 & 0.15 &4 & 0.63 & -126.1 & & \\
 \hline
551& PSF & 22.37 & 23.71 & & & & & 1.29 & 1.28\\
 &Bulge & 22.34 & 22.53 & 0.15 &4 & 0.70 & 42.2 & & \\
\hline
589& PSF & 21.50 & 22.33 & & & & & 1.29 & 1.63\\
 &Disk & 22.19 & 22.20 & 0.55 &1 & 0.47 & 31.6 & & \\
\hline
601& PSF & 21.33 & 22.20 & & & & & 1.29 & 1.62\\
 &Disk & 21.38 & 21.46 & 1.02 &1 & 0.58 & 56.2 & & \\
\hline
622& PSF & 20.55 & 21.59 & & & & & 6.73 & 1.12\\
 &Bulge & 21.65 & 21.24 & 0.50 &4 & 0.70 & 118.1 & & \\
\hline
634& PSF & 21.85 & 22.68 & & & & & 1.32 & 1.50\\
 &Disk & 22.17 & 22.38 & 0.76 &1 & 0.34 & 118.6 & & \\
\hline
645& PSF & 20.47 & 21.36 & & & & & 1.41 & 1.74\\
 &Bulge & 21.23 & 21.46 & 0.31 &4 & 0.89 & 77.7 & & \\
\hline
694& PSF & 20.41 & 21.57 & & & & & 1.20 & 1.63\\
 &Bulge & 23.07 & 23.85 & 0.78 &4 & 0.76 & -144.2 & & \\
\hline
720& PSF & 20.27 & 21.24 & & & & & 1.40 & 2.52\\
 &Bulge & 20.69 & 21.14 & 0.37 &4 & 0.87 & 50.3 & & \\
\hline
767& PSF & 21.59 & 22.20 & & & & & 1.18 & 1.31\\
 &Bulge & 21.16 & 21.26 & 1.48 &4 & 0.73 & 169.0 & & \\
\hline
772& PSF & 21.78 & 22.40 & & & & & 1.17 & 1.00\\
 &Bulge & 18.69 & 18.99 & 1.58 &4 & 0.88 & 88.8 & & \\
\hline
775& PSF & 19.73 & 21.02 & & & & & 1.19 & 2.14\\
 &Bulge & 19.58 & 19.69 & 0.18 &4 & 0.71 & -175.7 & & \\
 &Disk & 18.87 & 19.03 & 2.32 &1 & 0.78 & 139.7 & & \\
& {UVIS Trunc.} & 21.83 &  & 0.42 & 1 & 0.6 & 156.0 & & \\
& {Fourier} &  & -0.56 & 0.14 & 0.10 & -0.07 & 0.04 & & \\
& {Fourier} &  & 4.35 & 2.58 & 32.19 & 19.36 & 1.16 & & \\
& {Radial} &  &  & 0.33 & 1.08 & 0.40 & 152.5 & & \\
& {Fourier} &  & 0.53 & 0.33 & 0.17 & 0.04 & 0.02 & & \\
& {Fourier} &  & -161.59 & 4.08 & -43.21 & -3.69 & -2.25 & & \\
& {IR Trunc.} &  & 23.89 & 0.24 & 1 & 0.52 & 139.2 & & \\
& {Fourier} &  & -0.53 & 0.08 & 0.06 & -0.02 & 0.05 & & \\
& {Fourier} &  & 47.96 & 10.96 & 41.73 & 18.83 & 14.42 & & \\
& {Radial} &  &  & 0.88 & 1.38 & 0.52 & 136.5 & & \\
& {Fourier} &  & 0.56 & 0.11 & -0.07 & -0.03 & -0.07 & & \\
& {Fourier} &  & -125.58 & 12.36 & 5.47 & 0.66 & -16.27 & & \\
\hline
776& PSF & 20.34 & 21.40 & & & & & 0.43 & 0.09\\
 &Bulge & 19.33 & 19.32 & 0.46 &4 & 0.77 & 176.8 & & \\
 &Disk & 19.33 & 19.49 & 2.17 &1 & 0.35 & -156.8 & & \\
\hline\hline
 \end{tabular}
\end{table*}

\begin{table*}[t]
\movetableright=-0.35in
\setcounter{table}{1}
\caption{(Continued).}
\setlength\tabcolsep{6pt}
\label{tab:galfit2}
\begin{tabular}{c|crrrrrrrr}
\hline\hline
RMID & Comp. & ${\rm {mag}_{UVIS}}$ & ${\rm {mag}_{IR}}$ & r ($''$) & n & q & P. A. & ${\rm {r\chi^{2}}_{UVIS}}$ & ${\rm {r\chi^{2}}_{IR}}$\\
\hline		
\hline
779& PSF & 20.19 & 20.98 & & & & & 1.32 & 1.19\\
 &Disk & 20.38 & 20.84 & 0.75 &1 & 0.57 & -35.5 & & \\
\hline
781& PSF & 20.53 & 21.81 & & & & & 1.33 & 1.80\\
 &Bulge & 20.84 & 20.87 & 0.33 &4 & 0.45 & 68.4 & & \\
 &Disk & 20.86 & 21.27 & 1.01 &1 & 0.71 & -156.6 & & \\
\hline
782& PSF & 19.83 & 21.18 & & & & & 1.38 & 2.10\\
 &Bulge & 22.52 & 22.16 & 0.52 &4 & 0.28 & 48.2 & & \\
 &Disk & 20.21 & 20.32 & 1.63 &1 & 0.72 & -0.8 & & \\
\hline
790& PSF & 21.29 & 21.92 & & & & & 1.30 & 2.06\\
 &Bulge & 19.70 & 19.79 & 0.53 &4 & 0.75 & 122.1 & & \\
 &Disk & 21.48 & 20.95 & 0.47 &1 & 0.64 & 35.5 & & \\
& {UVIS Trunc.} & 25.47 &  & 0.71 & 1 & 0.50 & 121.7 & & \\
& {Radial} &  &  & 2.60 & 4.43 & 0.52 & 122.4 & & \\
& {IR Trunc.} &  & 25.22 & 0.91 & 1 & 0.49 & 122.1 & & \\
& {Radial} &  &  & 2.58 & 4.10 & 0.54 & 123.3 & & \\
\hline
840& PSF & 20.45 & 22.37 & & & & & 1.30 & 2.67\\
 &Bulge & 20.19 & 19.98 & 0.26 &4 & 0.86 & 36.1 & & \\
 &Disk & 19.82 & 19.97 & 2.95 &1 & 0.23 & 56.1 & & \\
\hline\hline
 \end{tabular}
 \tablecomments{$r$ is the effective radius of the S\'ersic component, $n$ is the S\'ersic index, $q$ is the ratio between the semi-minor axis and the semi-major axis, and P.A. is the position angle at the semi-major axis in degrees. The reduced ${\chi}^{2}$ is calculated from the image residual, as reported by {\tt GALFIT}. Magnitudes are reported in ST magnitude {(${\rm mag}_{\rm ST} = - 2.5 \log({F}_{\lambda}{\rm[erg\,s^{-1}\,cm^{-2}\,\AA^{-1}]}) - 21.1$)}, which is the default output from {\tt GALFIT}. {For RM775 and RM790, we include the best-fit parameters for the truncated disks in the UVIS and IR images: the magnitudes are the surface brightness at the break radius (mag/arcsec$^{2}$), and the best-fit parameters for the truncated radial profiles are listed in the order of the 1\% flux radius (softening length, in arcseconds), 99\% flux radius (break radius, in arcseconds), q, and P.A.. The truncated disk in RM775 is fitted with Fourier modes in both the disks and truncated radial profiles, and the best-fit Fourier amplitudes (first row) and phase angles (second row) are listed in the order of Fourier mode 1, 3, 4, 5, 6.}
 No extinction corrections are made for these magnitudes. The uncertainties of the {\tt GALFIT} results are discussed in Section \ref{sec:obs}.}
\end{table*}
 
\begin{table*}[t]
\movetableright=-0.5in
\caption{{Final photometry, Color, Luminosity, and Stellar mass}}
\label{tab:mass}
\begin{tabular}{cc|cccccccc}
\hline\hline
RMID & Bands & Comp & ${\rm {m}_{B}}$    & ${\rm {m}_{I/R}}$  & Color & $\log L_{\rm B}$          & $\log L_{\rm I/R}$        & $\log M_{*}$ & $\log M_{\rm *,CIGALE}$       \\
 &  & & (mag)   & (mag)& (mag) & ($L_{\rm \odot}$)          &  ($L_{\rm \odot}$)        &  ($M_{\rm \odot}$)   & ($M_{\rm \odot}$)        \\
\hline\hline
017 & B, R & Host & 19.66	&	18.78	&	0.88	&	11.49$\pm$0.10	&	11.07$\pm$0.10	&	11.06$\pm$0.35	&	10.91$\pm$0.31\\
& & Bulge & 19.86	&	19.03	&	0.84	&	11.40$\pm$0.10	&	10.98$\pm$0.10	&	10.92$\pm$0.35	&	10.77$\pm$0.31\\
& & Disk & 21.43	&	20.36	&	1.06	&	10.86$\pm$0.10	&	10.44$\pm$0.10	&	10.60$\pm$0.35	&	10.36$\pm$0.32\\
033 & B, R & Disk & 22.13 & 21.79 & 0.34 &10.76$\pm$0.10 & 10.34$\pm$0.10 & 9.83$\pm$0.35 & 9.88$\pm$0.28\\
101 & B, R & Bulge & 21.05 & 20.24 & 0.81 &10.92$\pm$0.10 & 10.50$\pm$0.10 & 10.42$\pm$0.35 & 10.25$\pm$0.30\\
160 & B, I & Bulge & 21.36 & 20.16 & 1.20 &10.70$\pm$0.10 & 10.15$\pm$0.10 & 9.94$\pm$0.27 & 9.95$\pm$0.30\\
177 & B, R & Disk & 20.86 & 20.38 & 0.48 &10.91$\pm$0.10 & 10.49$\pm$0.10 & 10.11$\pm$0.35 & 9.78$\pm$0.46\\
191 & B, R & Host & 20.61 & 19.76 & 0.85 &11.07$\pm$0.10 & 10.65$\pm$0.10 & 10.61$\pm$0.35 & 10.43$\pm$0.30\\
 &  & Bulge & 23.09 & 21.63 & 1.46 &10.32$\pm$0.10 & 9.90$\pm$0.10 & 10.43$\pm$0.35 & 10.00$\pm$0.36\\
 &  & Disk & 20.79 & 20.02 & 0.77 &10.97$\pm$0.10 & 10.55$\pm$0.10 & 10.43$\pm$0.35 & 10.28$\pm$0.30\\
229 & B, R & Host & 21.44 & 20.55 & 0.90 &10.82$\pm$0.10 & 10.40$\pm$0.10 & 10.41$\pm$0.35 & 10.23$\pm$0.31\\
 &  & Bulge & 24.40 & 22.67 & 1.73 &9.97$\pm$0.10 & 9.55$\pm$0.10 & 10.33$\pm$0.35 & 9.85$\pm$0.38\\
 &  & Disk & 21.58 & 20.77 & 0.81 &10.73$\pm$0.10 & 10.31$\pm$0.10 & 10.24$\pm$0.35 & 10.10$\pm$0.31\\
265 & B, R & Disk & 22.84 & 20.60 & 2.23 &11.27$\pm$0.10 & 10.85$\pm$0.10 & 12.10$\pm$0.35 & 11.39$\pm$0.46\\
267 & B, R & Bulge & 20.81 & 20.14 & 0.67 &11.22$\pm$0.10 & 10.80$\pm$0.10 & 10.59$\pm$0.35 & 10.49$\pm$0.31\\
272 & B, I & Host & 20.63 & 19.58 & 1.05 &10.62$\pm$0.10 & 10.07$\pm$0.10 & 9.76$\pm$0.27 & 9.81$\pm$0.29\\
 &  & Bulge & 21.67 & 20.38 & 1.29 &10.30$\pm$0.10 & 9.75$\pm$0.10 & 9.60$\pm$0.27 & 9.55$\pm$0.30\\
 &  & Disk & 21.14 & 20.28 & 0.86 &10.34$\pm$0.10 & 9.79$\pm$0.10 & 9.34$\pm$0.27 & 9.47$\pm$0.29\\
300 & B, I & Disk & 21.55 & 20.93 & 0.61 &11.00$\pm$0.10 & 10.45$\pm$0.10 & 9.83$\pm$0.27 & 10.06$\pm$0.28\\
301 & B, R & Host & 20.72 & 19.82 & 0.89 &11.27$\pm$0.10 & 10.85$\pm$0.10 & 10.85$\pm$0.35 & 10.66$\pm$0.31\\
 &  & Bulge & 22.19 & 20.91 & 1.28 &10.84$\pm$0.10 & 10.42$\pm$0.10 & 10.78$\pm$0.35 & 10.44$\pm$0.35\\
 &  & Disk & 21.21 & 20.47 & 0.74 &11.01$\pm$0.10 & 10.59$\pm$0.10 & 10.44$\pm$0.35 & 10.32$\pm$0.31\\
305 & B, R & Bulge & 20.88 & 20.01 & 0.87 &11.16$\pm$0.10 & 10.73$\pm$0.10 & 10.72$\pm$0.35 & 10.55$\pm$0.31\\
316 & B, I & Bulge & 20.12 & 19.68 & 0.44 &11.55$\pm$0.10 & 11.00$\pm$0.10 & 10.25$\pm$0.27 & 10.57$\pm$0.26\\
320 & B, I & Host & 19.91 & 18.70 & 1.21 &10.98$\pm$0.10 & 10.43$\pm$0.10 & 10.23$\pm$0.27 & 10.21$\pm$0.30\\
 &  & Bulge & 22.01 & 20.61 & 1.40 &10.21$\pm$0.10 & 9.66$\pm$0.10 & 9.60$\pm$0.27 & 9.52$\pm$0.30\\
 &  & Disk & 20.06 & 18.90 & 1.16 &10.90$\pm$0.10 & 10.34$\pm$0.10 & 10.11$\pm$0.27 & 10.11$\pm$0.29\\
338 & B, R & Disk & 20.89 & 20.27 & 0.62 &10.81$\pm$0.10 & 10.39$\pm$0.10 & 10.14$\pm$0.35 & 10.07$\pm$0.30\\
371 & B, R & Bulge & 20.70 & 19.87 & 0.83 &11.09$\pm$0.10 & 10.67$\pm$0.10 & 10.62$\pm$0.35 & 10.45$\pm$0.31\\
377 & B, I & Host & 20.54	&	19.24	&	1.30	&	11.00$\pm$0.10	&	10.45$\pm$0.10	&	10.32$\pm$0.27	&	10.23$\pm$0.30\\
& & Bulge & 21.16	&	19.80	&	1.36	&	10.78$\pm$0.10	&	10.23$\pm$0.10	&	10.14$\pm$0.27	&	10.06$\pm$0.30\\
& & Disk & 21.45	&	20.37	&	1.08	&	10.55$\pm$0.10	&	10.00$\pm$0.10	&	9.71$\pm$0.27	&	9.74$\pm$0.30\\
392 & B, R & Bulge & 21.76 & 20.98 & 0.77 &11.26$\pm$0.10 & 10.84$\pm$0.10 & 10.73$\pm$0.35 & 10.58$\pm$0.32\\
457 & B, R & Bulge & 22.04 & 20.89 & 1.15 &10.94$\pm$0.10 & 10.39$\pm$0.10 & 10.15$\pm$0.27 & 10.15$\pm$0.31\\
519 & B, R & Bulge & 23.30	&	22.57	&	0.73	&	10.18$\pm$0.10	&	9.76$\pm$0.10	&	9.61$\pm$0.35	&	9.50$\pm$0.31\\
551 & B, I & Bulge & 22.20 & 21.12 & 1.08 &10.98$\pm$0.10 & 10.43$\pm$0.10 & 10.14$\pm$0.27 & 10.14$\pm$0.31\\
589 & B, R & Disk & 22.05	&	21.13	&	0.92	&	11.08$\pm$0.10	&	10.66$\pm$0.10	&	10.69$\pm$0.35	&	10.44$\pm$0.32\\
601 & B, I & Disk & 21.15 & 20.07 & 1.07 &11.36$\pm$0.10 & 10.81$\pm$0.10 & 10.52$\pm$0.27 & 10.52$\pm$0.31\\
622 & B, R & Bulge & 21.25 & 20.33 & 0.91 &11.11$\pm$0.10 & 10.69$\pm$0.10 & 10.71$\pm$0.35 & 10.54$\pm$0.31\\
634 & B, I & Disk & 21.81 & 20.97 & 0.84 &10.99$\pm$0.10 & 10.44$\pm$0.10 & 9.98$\pm$0.27 & 10.09$\pm$0.30\\
645 & B, R & Bulge & 21.14 & 20.52 & 0.62 &10.84$\pm$0.10 & 10.42$\pm$0.10 & 10.16$\pm$0.35 & 10.08$\pm$0.30\\
694 & B, R & Bulge & 23.10 & 22.81 & 0.29 &10.05$\pm$0.10 & 9.63$\pm$0.10 & 9.07$\pm$0.35 & 9.10$\pm$0.25\\
720 & B, R & Bulge & 20.62 & 20.14 & 0.48 &10.97$\pm$0.10 & 10.55$\pm$0.10 & 10.17$\pm$0.35 & 10.14$\pm$0.29\\
767 & B, R & Bulge & 21.02 & 20.32 & 0.71 &11.03$\pm$0.10 & 10.61$\pm$0.10 & 10.44$\pm$0.35 & 10.31$\pm$0.31\\
772 & B, I & Bulge & 18.88 & 17.84 & 1.04 &11.26$\pm$0.10 & 10.71$\pm$0.10 & 10.39$\pm$0.27 & 10.42$\pm$0.29\\
775 & B, I & Host & 18.72 & 17.46 & 1.25 &11.05$\pm$0.10 & 10.50$\pm$0.10 & 10.34$\pm$0.27 & 10.27$\pm$0.29\\
 &  & Bulge & 19.84 & 18.64 & 1.20 &10.58$\pm$0.10 & 10.03$\pm$0.10 & 9.83$\pm$0.27 & 9.82$\pm$0.29\\
 &  & Disk & 19.10 & 17.97 & 1.13 &10.85$\pm$0.10 & 10.30$\pm$0.10 & 10.04$\pm$0.27 & 10.08$\pm$0.29\\
\hline\hline
\end{tabular}
\end{table*}

\begin{table*}[t]
\movetableright=-0.5in
\setcounter{table}{2}
\caption{(continued).}
\begin{tabular}{cc|cccccccc}
\hline\hline
RMID & Bands & Comp & ${\rm {m}_{B}}$    & ${\rm {m}_{I/R}}$  & Color & $\log L_{\rm B}$          & $\log L_{\rm I/R}$        & $\log M_{*}$ & $\log M_{\rm *,CIGALE}$       \\
 &  & & (mag)   & (mag)& (mag) & ($L_{\rm \odot}$)          &  ($L_{\rm \odot}$)        &  ($M_{\rm \odot}$)   & ($M_{\rm \odot}$)        \\
\hline\hline
776 & B, I & Host & 18.93	&	17.62	&	1.31	&	10.61$\pm$0.10	&	10.06$\pm$0.10	&	9.94	$\pm$0.27	&	9.87	$\pm$0.30\\
& & Bulge & 19.73	&	18.31	&	1.41	&	10.34$\pm$0.10	&	9.79$\pm$0.10	&	9.73$\pm$0.27	&	9.62$\pm$0.30\\
& & Disk & 19.63	&	18.44	&	1.19	&	10.29$\pm$0.10	&	9.74$\pm$0.10	&	9.53$\pm$0.27	&	9.52$\pm$0.29\\
779 & B, I & Disk & 20.64 & 19.65 & 0.99 &10.06$\pm$0.10 & 9.51$\pm$0.10 & 9.16$\pm$0.27 & 9.18$\pm$0.28\\
781 & B, I & Host & 20.29 & 19.14 & 1.15 &10.79$\pm$0.10 & 10.24$\pm$0.10 & 10.00$\pm$0.27 & 9.98$\pm$0.29\\
 &  & Bulge & 21.05 & 19.76 & 1.30 &10.55$\pm$0.10 & 10.00$\pm$0.10 & 9.86$\pm$0.27 & 9.79$\pm$0.30\\
 &  & Disk & 21.06 & 20.06 & 1.01 &10.43$\pm$0.10 & 9.88$\pm$0.10 & 9.54$\pm$0.27 & 9.54$\pm$0.29\\
782 & B, I & Host & 20.08 & 18.94 & 1.14 &11.20$\pm$0.10 & 10.65$\pm$0.10 & 10.40$\pm$0.27 & 10.39$\pm$0.30\\
 &  & Bulge & 22.52 & 20.98 & 1.54 &10.38$\pm$0.10 & 9.83$\pm$0.10 & 9.87$\pm$0.27 & 9.67$\pm$0.31\\
 &  & Disk & 20.22 & 19.14 & 1.08 &11.12$\pm$0.10 & 10.56$\pm$0.10 & 10.28$\pm$0.27 & 10.31$\pm$0.30\\
790 & B, I & Host & 19.76 & 18.41 & 1.35 &10.99$\pm$0.10 & 10.43$\pm$0.10 & 10.34$\pm$0.27 & 10.26$\pm$0.30\\
 &  & Bulge & 19.96 & 18.65 & 1.31 &10.89$\pm$0.10 & 10.34$\pm$0.10 & 10.21$\pm$0.27 & 10.10$\pm$0.30\\
 &  & Disk & 21.86 & 19.94 & 1.91 &10.37$\pm$0.10 & 9.82$\pm$0.10 & 10.12$\pm$0.27 & 9.76$\pm$0.33\\
840 & B, I & Host & 19.44 & 18.14 & 1.30 &11.12$\pm$0.10 & 10.57$\pm$0.10 & 10.44$\pm$0.27 & 10.38$\pm$0.30\\
 &  & Bulge & 20.41 & 18.95 & 1.46 &10.79$\pm$0.10 & 10.24$\pm$0.10 & 10.22$\pm$0.27 & 10.12$\pm$0.31\\
 &  & Disk & 20.01 & 18.86 & 1.15 &10.83$\pm$0.10 & 10.28$\pm$0.10 & 10.04$\pm$0.27 & 10.04$\pm$0.29\\
\hline\hline
\end{tabular}
\tablecomments{Magnitudes are reported in AB magnitudes \citep{Oke_Gunn_1982}, and color refers to either $B-I$ or $B-R$. The last column lists the stellar masses estimated with {\tt CIGALE} to compare with our fiducial stellar masses. }
\end{table*}

\subsection{Host Galaxy/Bulge Masses}
Following the approach in \cite{Li_etal_2021}, we convert the UVIS/IR photometry to rest-frame $B$ and $I/R$ band and estimate the host/bulge stellar masses with the color-$M_{\rm *}/L$  relations (CMLR) from \cite{Into_etal_2013} and {{\tt CIGALE} \citep{Boquien_etal_2019}}. First, we correct for Galactic extinction using the recalibrated \cite{Schlegel_etal_1998} dust map and reddening from \cite{Schlafly_etal_2011}. We then fit the extinction-corrected HST photometry with {\tt CIGALE} to derive k-corrections and color transformations between the HST filters and the Johnson$-$Cousins filters. {{\tt CIGALE} is a spectral energy distribution (SED) fitting code that can model galaxy and AGN emission from multiwavelength photometry. We set up a simple {\tt CIGALE} model that includes basic stellar population synthesis models \citep{Maraston_2005}, a initial mass function \citep{Kroupa_2001}, a dust attenuation model \citep{Calzetti_etal_2000,Leitherer_etal_2002}, and a delayed star formation history with optional starburst. We do not include the AGN model for modeling the quasar-subtracted photometry.} The UVIS filters are converted to $B$-band magnitudes, and the IR filters are converted to $I$-band and $R$-band filters depending on the source redshift (F110W to $I$-band at $z<0.4$ and $R$-band $z>0.4$; F140W to $I$-band at $z<0.7$ and $R$-band at $z>0.7$). 

{We estimate the host and bulge stellar masses with the CMLR for dusty galaxy models from \cite{Into_etal_2013} using the rest-frame photometry and their uncertainties. {\tt CIGALE} fits provide the k-corrected photometry, from which we estimate stellar mass with the CMLR relation, and a stellar mass from the best-fit SED model.} For RM177 ({with two UVIS bands from a separate HST program}), we include both the F606W and F814W bands for the CIGALE fitting but only use the F606W band (rest-frame $R$-band) for the CMLR stellar mass estimation. {The CMLR stellar-mass uncertainties are propagated directly from the photometry uncertainties, and the CIGALE stellar-mass uncertainties are estimated from the SED modeling. Both CMLR and CIGALE uncertainties are consistently around 0.3\,dex, which is typical for stellar mass estimation from two-band photometry.}

\begin{figure*}[t]
\centering
\includegraphics[width=0.46\textwidth]{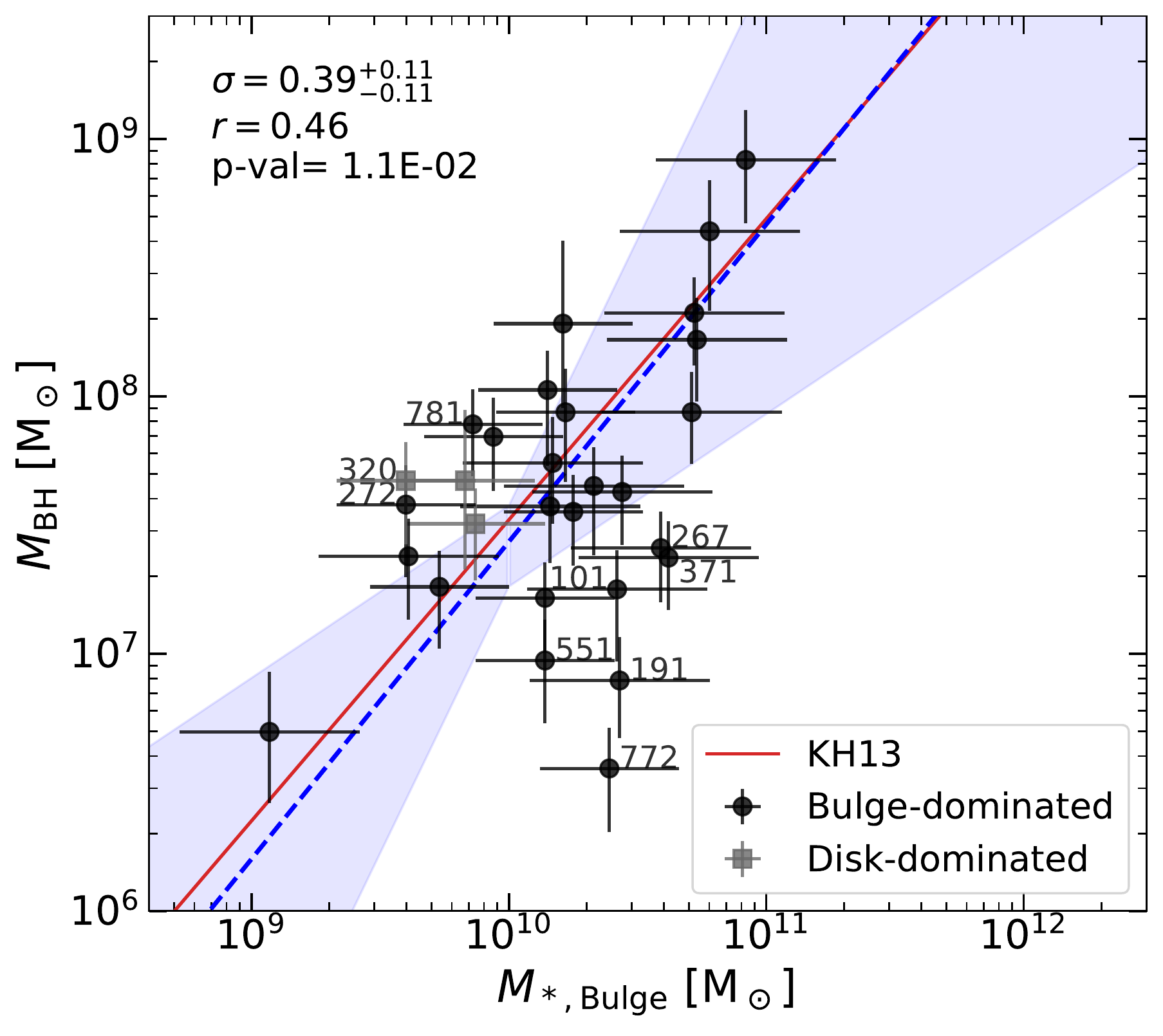}
\includegraphics[width=0.46\textwidth]{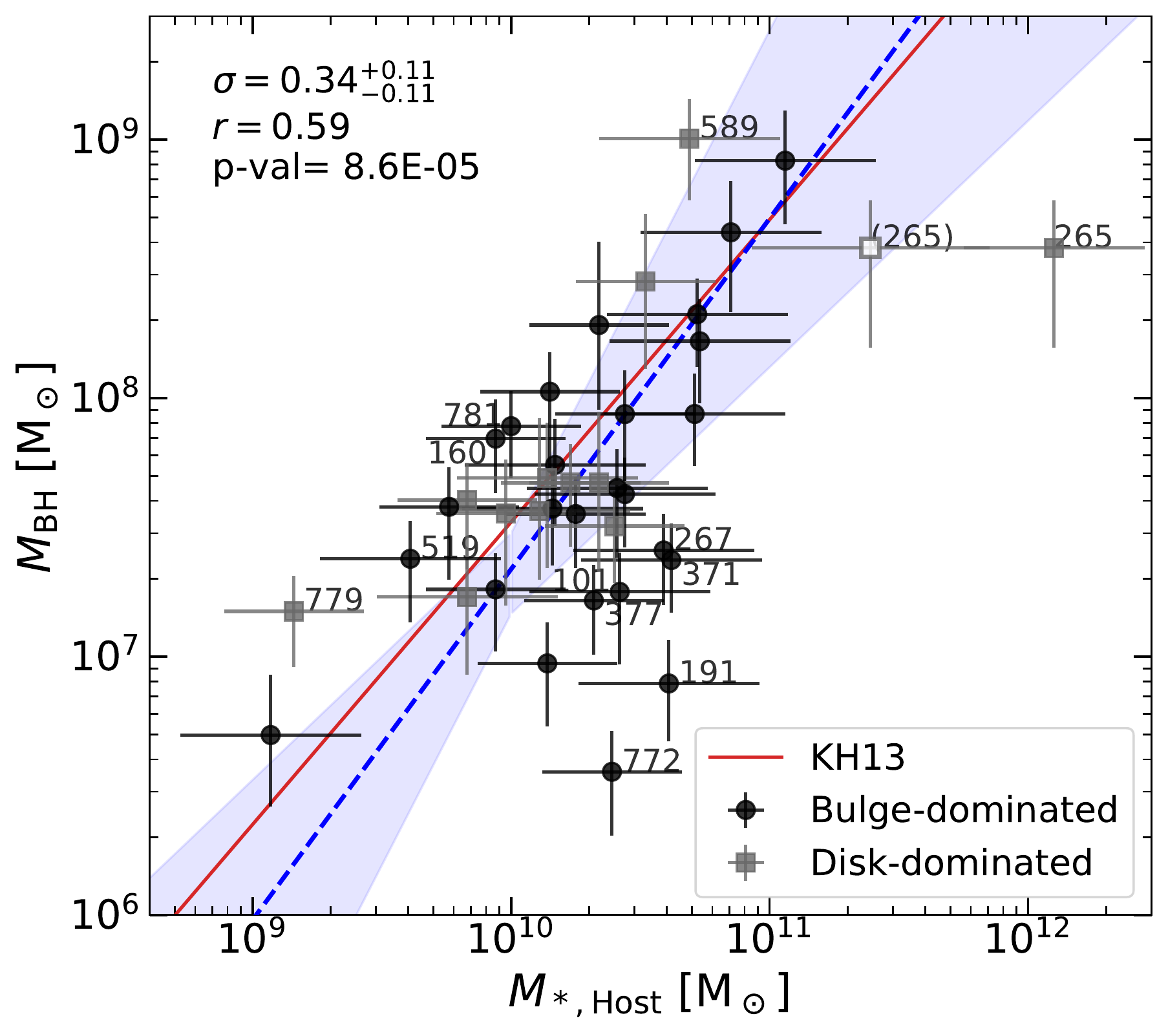}\\
\caption{BH mass as functions of bulge stellar mass (left) and total stellar mass (right) of our sample (black circles for bulge-dominated sources, gray squares for disk-dominated sources, and the open gray square for the {\tt CIGALE} stellar mass of RM265). The blue dashed lines (and the gray shaded areas) are the best-fit relation (1$\sigma$ range) of our sample. The red solid lines are the \cite{Kormendy_Ho_2013} local $M_{\rm BH}-M_{\rm *, bulge}$ relation. The Pearson-r coefficient, p-value, and intrinsic scatter of the relations are labeled in the top-left corner of each panel.} 
\label{fig:Ms_Mgal}
\end{figure*}

The final Galactic-extinction corrected, k-corrected, band-converted magnitudes, and the host/bulge stellar masses are tabulated in Table \ref{tab:mass}. {We adopt the CMLR stellar masses as our nominal host/bulge stellar masses.} The best-fit stellar masses from {\tt CIGALE} are also reported for comparison, {which are generally consistent with those estimated from the CMLR. The only exception is RM265. The color derived for RM265 from {\tt CIGALE} is unusually red, which led to a large, likely unphysical, host stellar mass ($>10^{12} M_{\odot}$) using the CMLR. However, the typical $B-R$ color is roughly $0.3<(B-R)<1.3$, derived from all galaxy types in the Kinney-Calzetti Spectral Atlas \citep{Calzetti_etal_1994,Kinney_etal_1996}. If we assume a red color of 1.3 and adopt the $R$-band luminosity for the CMLR, the host stellar mass for RM265 is ${\rm log}(M_{\rm *})=11.23$, which is consistent with the stellar mass derived from {\tt CIGALE}.} {We show both the CMLR and {\tt CIGALE} masses for RM265 in Figures \ref{fig:Ms_Mgal}, \ref{fig:fhost}, and \ref{fig:redshift}, and use the more physical {\tt CIGALE} mass (for RM265 only) when fitting the BH scaling relations and their redshift evolution in our analysis.}

\section{Results} \label{sec:results}

\subsection{Host Properties}

At $z>0.2$, it becomes challenging to perform bulge/disk decomposition due to limited spatial resolution, even with HST. Our {\tt GALFIT} analysis shows 16 (out of 38) quasars are best-fitted by the PSF+bulge model, nine quasars are best-fitted by the PSF+disk model, and 13 quasars are decomposed into PSF+bulge+disk models. In addition, 26 hosts are bulge-dominated, i.e., $M_{\rm *, bulge}>M_{\rm *, disk}$, and 12 hosts are disk-dominated. A best-fit profile of $n=4$ ($n=1$) in our analysis does not necessarily mean the host galaxy is an elliptical (spiral) galaxy; the S\'ersic index is fixed to $n=1$ or $n=4$ to ensure the quasar/host decomposition is robust and not to provide rigorous classifications of host morphology. {In fact, the majority of local elliptical galaxies are not well-described by single S\'ersic components \citep[e.g.,][]{Huang_etal_2013}, and exponential profiles do not always indicate the presence of disks.} 

The structural parameters (ellipticity, S\'ersic index, effective radius) of the bulge/disk-dominated sources in our sample are broadly consistent with the statistical distributions from {$\sim$\,2500, $i$-mag$<$22} SDSS quasar hosts observed by the Hyper Suprime-Cam (HSC) on the Subaru telescope \citep{LiJunyao_etal_2021a}. {When we allow the S\'ersic index to vary in the {\tt GALFIT} fitting, the median (minimum, maximum) S\'ersic index of our two-component model is 2.0 (0.6/7.0), similar to the distribution in \cite{LiJunyao_etal_2021a}. There are more disk-like ($n<2$) hosts in the SDSS-HSC sample, but roughly equal numbers of bulge-like and disk-like hosts (S\'ersic indices above and below 2) in our sample.} The size and ellipticity of our quasar hosts are also similar to the SDSS-HSC sample. The median (16\%, 84\% percentiles) effective radius is 0$''$.68 (0$''$.35/0$''$.92), and the median (16\%/84\% percentiles) ellipticity ($1-q$) is 0.28 (0.11/0.41) for our sample.

\begin{figure}[t]
\centering
\includegraphics[width=0.45\textwidth]{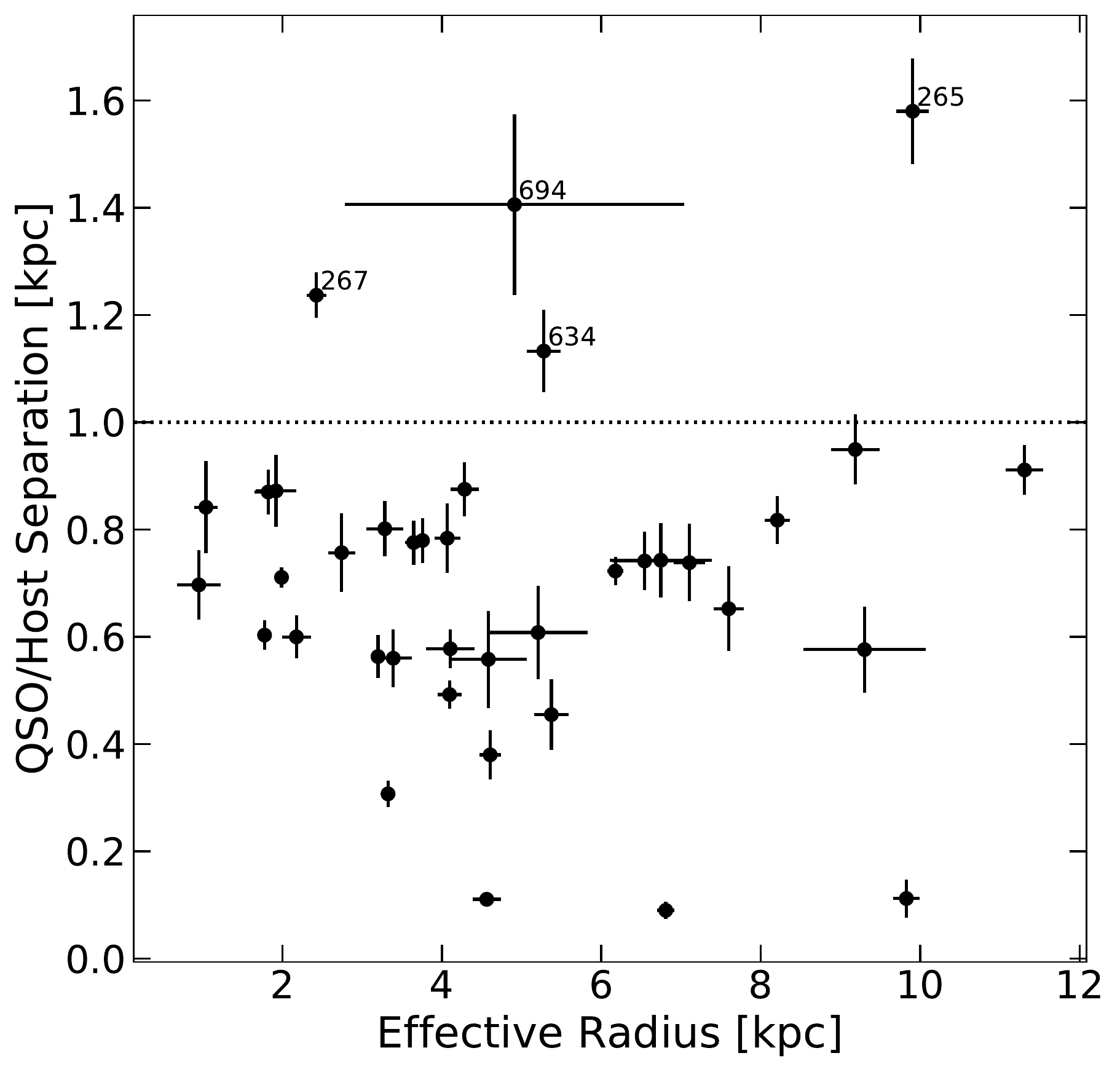}\\
\caption{{Projected physical offset between the quasar position and host centroid in the IR images as a function of the host effective radius. The uncertainties are estimated from {\tt GALFIT} (quasar position and effective radius) and the quasar-subtracted host images (centroid position of the host). The error bars are inflated by a factor of 10 for clarity, as the {\tt GALFIT} uncertainties are small and likely underestimated (median uncertainties are $\sim$0.02\,kpc for the quasar/host separation and $\sim$0.005\,kpc for the effective radius).}}
\label{fig:offset}
\end{figure}

{We also examine the offset between the quasar position and the host centroid in the IR images, where the centroid of the host galaxy is better constrained than in the UVIS band. Off-centered AGN/quasars may indicate on-going galaxy mergers or recoiling SMBHs from binary SMBH coalescence \citep[][]{Loeb_2007, Comerford_Greene_2014}. Figure \ref{fig:offset} shows that most (34/38) of the quasars are located within $<$\,1\,kpc of the host galaxy center. The four sources with significant offsets ($>$\,1\,kpc, RM265, RM267, RM634, RM645; see the images and {\tt GALFIT} models in full Figure \ref{fig:galfit_example} figure set online) show signs of galaxy interaction or mergers, which would complicate the centroid measurements of the host galaxy. These results suggest $z<1$ quasars are well centered within $\sim 1\,$kpc of the host centroid, consistent with the findings using alternative approaches \citep{Shen_etal_2019_vodka}.}

While studies of local AGNs have demonstrated that BH properties mainly correlate with the bulge and not the entire host \citep[e.g., ][]{Kormendy_Ho_2013}, studies at higher redshift are often limited to the BH$-$host relations when bulge/disk decomposition is difficult or impossible \citep[e.g.,][]{Jahnke_etal_2009, Merloni_etal_2010}. In this work, we present both the BH$-$bulge and the BH$-$host relations in our sample, where $M_{\rm *, bulge}$ and $M_{\rm *, host}$ refer to the bulge-only and total host stellar mass, respectively. We include all sources in the $M_{\rm BH}-M_{\rm *, host}$ relation and exclude {the disk-only (PSF+disk)} objects in the $M_{\rm BH}-M_{\rm *, bulge}$ relation. When comparing with earlier work, we examine their bulge/disk decomposition assumptions and place the comparison on an equal footing, i.e., including bulge-dominated or bulge/disk decomposed sources only in the $M_{\rm BH}-M_{\rm *, bulge}$ relation, and including all sources in the $M_{\rm BH}-M_{\rm *, host}$ relation.

\begin{figure*}[t]
\centering
\includegraphics[width=0.48\textwidth]{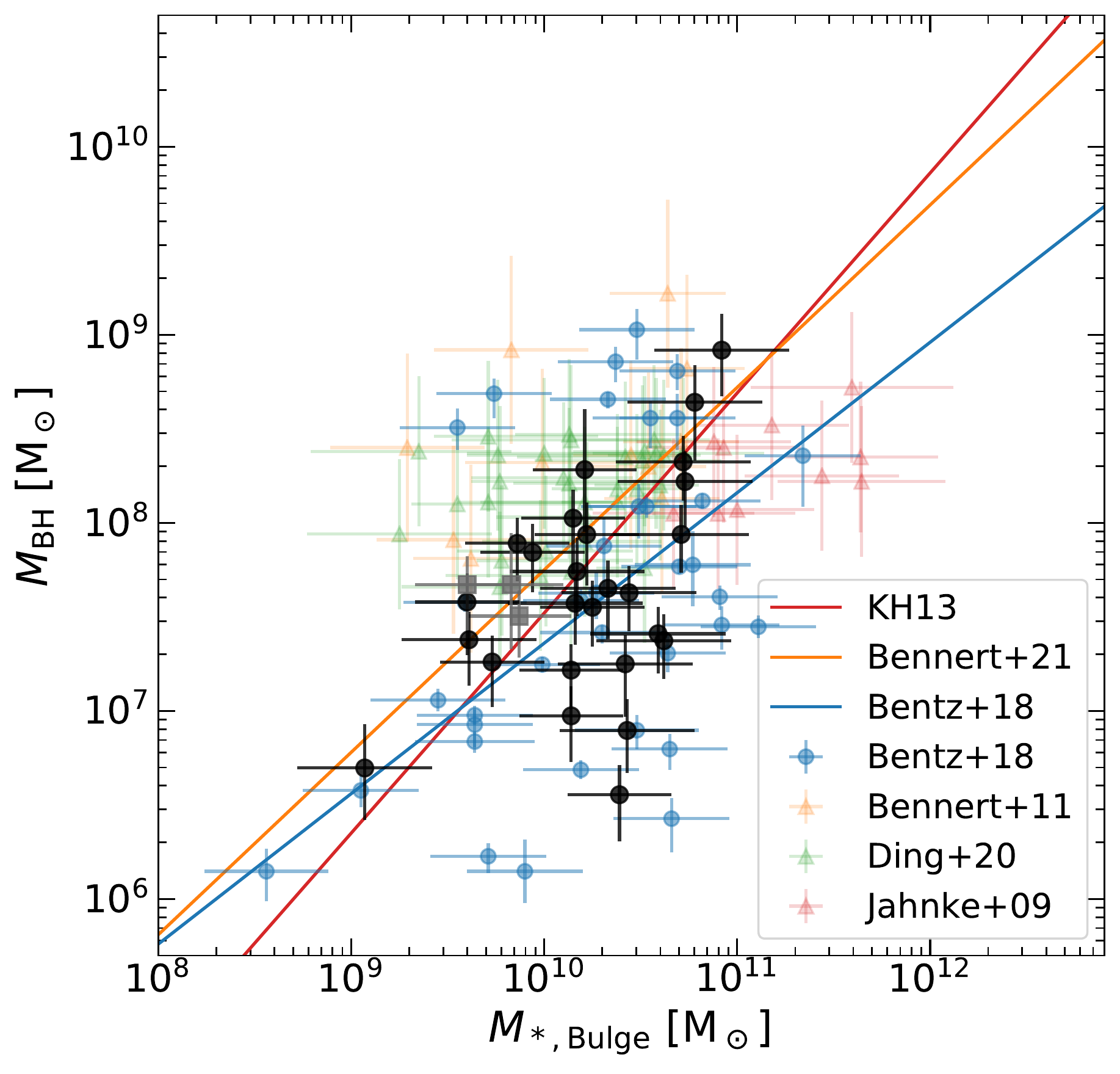}
\includegraphics[width=0.48\textwidth]{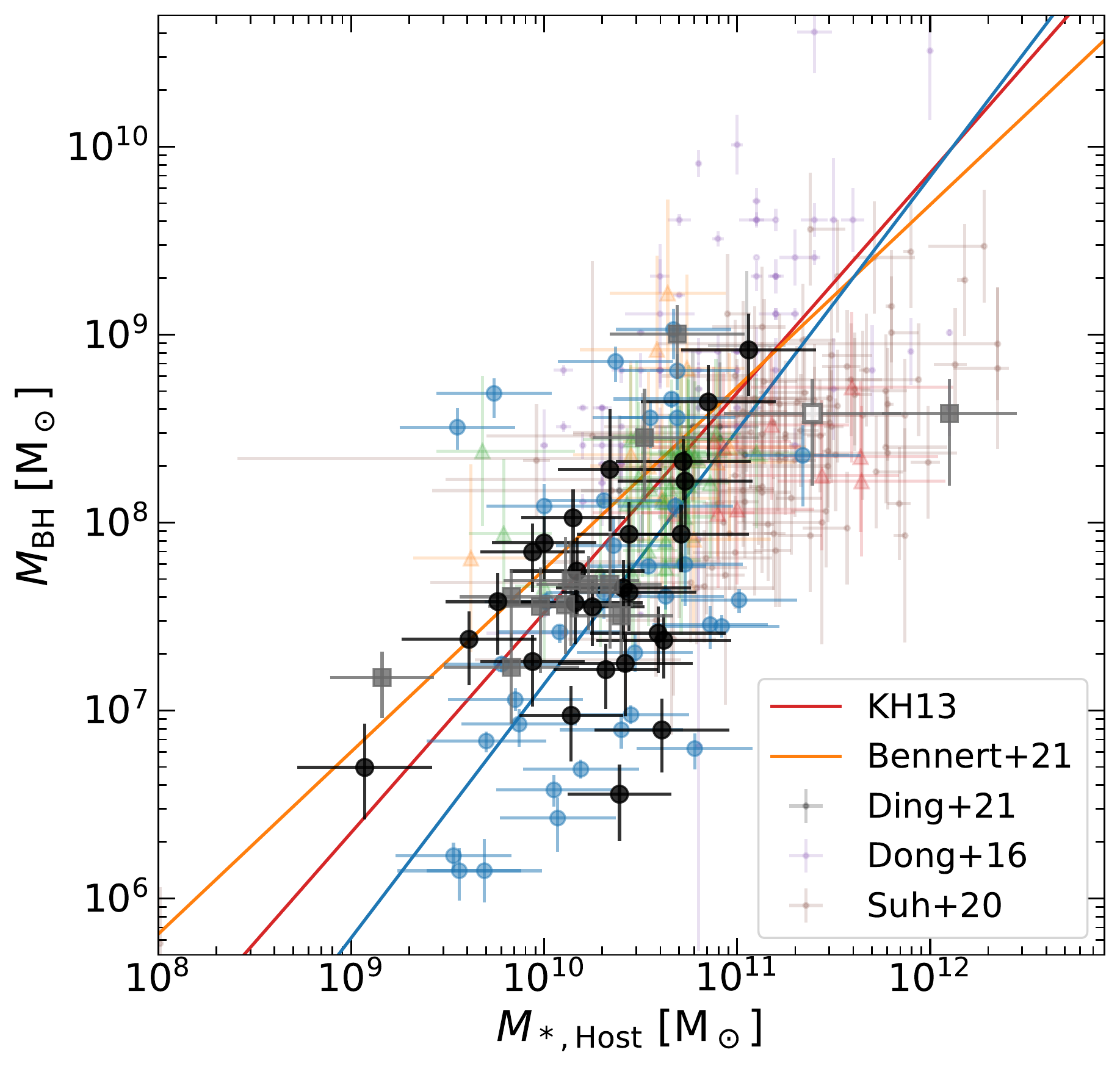}
\caption{Comparison with literature samples. {Our sample is shown in black circles (bulge-dominated) and gray squares (disk-dominated), and the open gray square is the {\tt CIGALE} stellar mass of RM265.} The blue points are the local-RM sample from \cite{Bentz_etal_2018}. {The triangle symbols show the high redshift samples with HST imaging, orange: \citet[][median redshift $z_{\rm med}=1.2$]{Bennert_etal_2011}, green: \citet[][$z_{\rm med}=1.3$]{Jahnke_etal_2009}, and red: \citet[][$z_{\rm med}=1.5$]{Ding_etal_2020}, and the dot symbols indicate the high redshift samples with SED fitting, purple: \citet[][$z_{\rm med}=1.7$]{Ding_etal_2021}, brown: \citet[][$z_{\rm med}=1.1$]{Dong_etal_2016}, pink: \citet[][$z_{\rm med}=1.6$]{Suh_etal_2020}.} The solid lines show the best-fit relations of the local quiescent galaxy \citep[red,][]{Kormendy_Ho_2013}, active galaxy \citep[orange,][]{Bennert_etal_2021}, and RM AGN \citep[blue,][]{Bentz_etal_2018}. } 
\label{fig:Ms_Mgal_lit}
\end{figure*}

\subsection{Comparison with earlier work}

Figure \ref{fig:Ms_Mgal_lit} shows the $M_{\rm BH}-M_{\rm *, host}$ and $M_{\rm BH}-M_{\rm *, bulge}$ relations of our sample and several local and higher-redshift samples. High-resolution HST imaging has been used to investigate the AGN host galaxies at $z>0.2$ \citep[e.g., ][]{Jahnke_etal_2009, Bennert_etal_2011, Bentz_etal_2018, Ding_etal_2020}. Similar to our study, bulge/disk decomposition is only possible for a small subset of these non-local samples. We follow the approach of \cite{Jahnke_etal_2009, Bennert_etal_2011, Bentz_etal_2018} and assume $M_{\rm *, bulge}\approx M_{\rm *, host}$ when there is no evidence of additional components, which differs from \cite{Ding_etal_2020} that estimated $M_{\rm *, bulge}$ by assigning bulge/total ratios depending on the S\'ersic indices of the host profile.

At even higher redshift (e.g., $z\gtrsim$1.5), host stellar masses can be obtained by SED fitting \citep[e.g.,][]{Merloni_etal_2010, Dong_etal_2016, Suh_etal_2020} or imaging analysis of lensed quasars \citep{Peng_etal_2006b, Ding_etal_2021}. SED fitting with wide wavelength coverage can provide better color information for estimating the stellar masses (compared to using only two HST bands), and large samples can be studied simultaneously in multiwavelength fields. However, it is impossible to distinguish between the bulge and disk components through SED fitting. $M_{\rm *, host}$ can also be measured from the reconstructed images of strongly lensed quasars up to $z\sim3$. Our sample is generally consistent with the $M_{\rm BH}-M_{\rm *, bulge}$ and $M_{\rm BH}-M_{\rm *, host}$ relations of these intermediate-to-high redshift samples.

Our sample is the only uniformly-selected {(i.e., selected based on a flux limit)} AGN sample with RM-based BH masses to study the BH scaling relations beyond the local Universe ($z>0.1$). {The RM masses in \cite{Grier_etal_2017} are consistently calibrated to the BH$-$host relations in quiescent local galaxies in \cite{Kormendy_Ho_2013} using the virial factor from \cite{woo_etal_2015}.} BH masses in all the comparison samples, except for \citet{Bentz_etal_2018}, are derived from the SE method, which is less reliable than RM masses. {The SE method relies on a ``tight'' $R-L$ relation to estimate BLR sizes based on quasar luminosities; however, recent studies \citep{Du_etal_2016, FonsecaAlvarez_etal_2020} have shown the local $R-L$ relation is biased towards the local AGN sample and could be overestimating SE BH masses by as much as $\sim$\,0.3\,dex when applying to the general quasar population.} Finally, the SE method is calibrated to local quiescent galaxies or local RM AGNs, and different virial factors may be used for different samples or broad-line species. When comparing to samples from the literature, we rescale all $M_{\rm BH}$ values using $f=4.47$ as in \cite{Grier_etal_2017}, even for the SE masses. 

{ We also compare our results with local baseline samples from the literature, including the quiescent galaxies \citep[mainly ellipticals,][]{Kormendy_Ho_2013}, active galaxies \citep{Bennert_etal_2021}, and RM AGNs \citep{Bentz_etal_2018}. The $M_{\rm BH}-M_{\rm *, bulge}$ and $M_{\rm BH}-M_{\rm *, host}$ relations of the three local samples and our best-fit relations are consistent in slope and intercepts within uncertainties. {However, \cite{Reines_Volonterni_2015} find AGN hosts follow a similar slope as local quiescent galaxies but are an order of magnitude lower in normalization for the $M_{\rm BH}-M_{\rm *, host}$ relation. They suggested the difference in normalization may be due to AGN activity or galaxy morphology \citep[which is also shown in][]{Greene_etal_2020}. These results may appear contradictory at first glance; however, it is difficult to provide a straightforward comparison since these studies adopt different stellar-mass estimation methods. \cite{Bentz_etal_2018} also observed a difference in normalization using the \cite{Bell_deJong_2001} CMLR, but not when they use the \cite{Into_etal_2013} CMLR, which is the same CMLR adopted in this work. Due to different assumptions in the CMLR relations, we do not compare the \cite{Reines_Volonterni_2015} relation, {which uses the \cite{Zibetti_etal_2009} CMLR,} with our results directly.}}

\cite{Sijacki_etal_2015} and \cite{Mutlu-Pakdil_etal_2018} studied the $M_{\rm BH}-M_{\rm *, bulge}$ and $M_{\rm BH}-M_{\rm *, host}$ relations in the Illustris simulation. \cite{Sijacki_etal_2015} found that, at $z\sim0$, the $M_{\rm BH}-M_{\rm *, bulge}$ relation is tight at the high-BH/galaxy mass end, but scatter increases below $M_{\rm BH}\sim10^{8}M_{\odot}$, similar to the general trend of our sample. Their bulge mass is defined by the total stellar mass within the stellar half-mass radius, and not by morphology or kinematics. The difference in scatter in the high/low mass end might suggest different evolutionary paths or feedback mechanisms for establishing the BH scaling relations. Local studies of BH scaling relations also found that late-type galaxies follow a similar slope in the BH scaling relations, but at a lower normalization than early-type galaxies  \citep{Reines_Volonterni_2015, Greene_etal_2020, Zhao_etal_2021}. In addition, there is no strong evolution in the $M_{\rm BH}-M_{\rm *, bulge}$ relation up to $z\sim1$ in the Illustris simulation. \cite{Mutlu-Pakdil_etal_2018} studied the $M_{\rm BH}-M_{\rm *, host}$ relations in the Illustris simulation to provide a better comparison for high redshift observations, and reported that the $M_{\rm BH}-M_{\rm *, bulge}$ and $M_{\rm BH}-M_{\rm *, host}$ relations are generally consistent with each other up to $z\sim1$. \cite{Volonteri_etal_2016} studied the $M_{\rm BH}-M_{\rm *, host}$ and $M_{\rm BH}-M_{\rm *, bulge}$ relations in the Horizon-AGN simulation. By identifying classical bulges in their simulation through kinematics and bulge/disk decomposition, they reproduced the tight $M_{\rm BH}-M_{\rm *, bulge}$ relation of classic bulges from \cite{Kormendy_Ho_2013}. Other simulations, e.g., MassiveBlack-II \citep{Khandai_etal_2015}, generally produce similar trends in BH scaling relations at $z<1$. \cite{Habouzit_etal_2021} performed a systematic analysis on the evolution of $M_{\rm BH}-M_{\rm *}$ relations in cosmological simulations of Illustris, TNG 100, TNG 300, Horizon-AGN, EAGLE and SIMBA. They find that the median/mean $M_{\rm BH}-M_{\rm *}$ relations at $0<z<1$ are in general agreement with observational data and there is little evolution with redshift. {The observed tight correlation between BH accretion rate and star formation at $0.5<z<3$ indicates the growth of BH and host galaxies are in sync, and the BH scaling relations should not have strong redshift dependence \citep{Yang_etal_2019}.} However, the scatter in $M_{\rm BH}-M_{\rm *}$ relations differs in these simulations, which mainly depends on the implemented sub-grid physics in the simulations, e.g., the strength and efficiency of supernova and AGN feedback.

\begin{figure*}[t]
\centering
\includegraphics[width=0.48\textwidth]{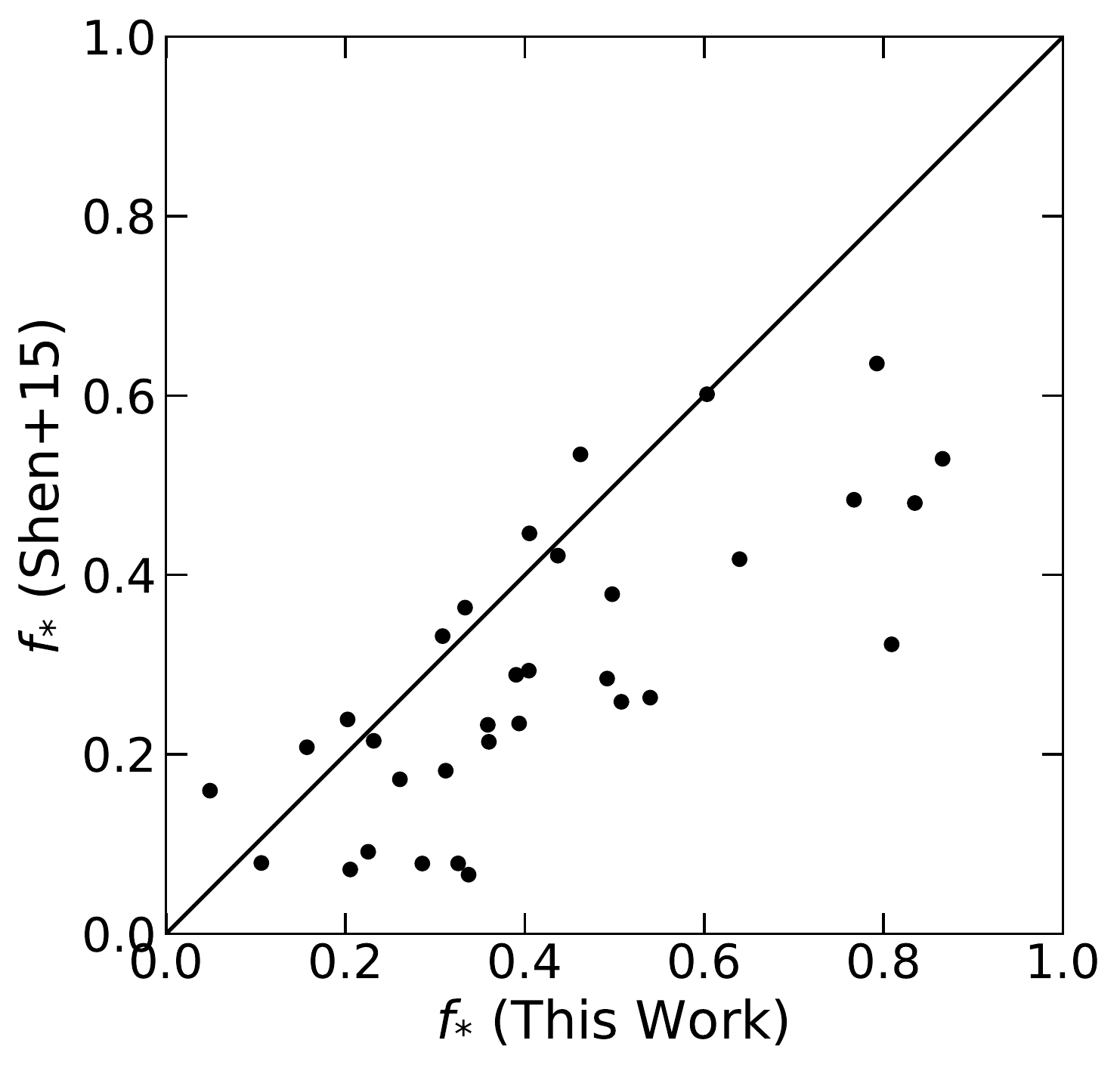}
\includegraphics[width=0.48\textwidth]{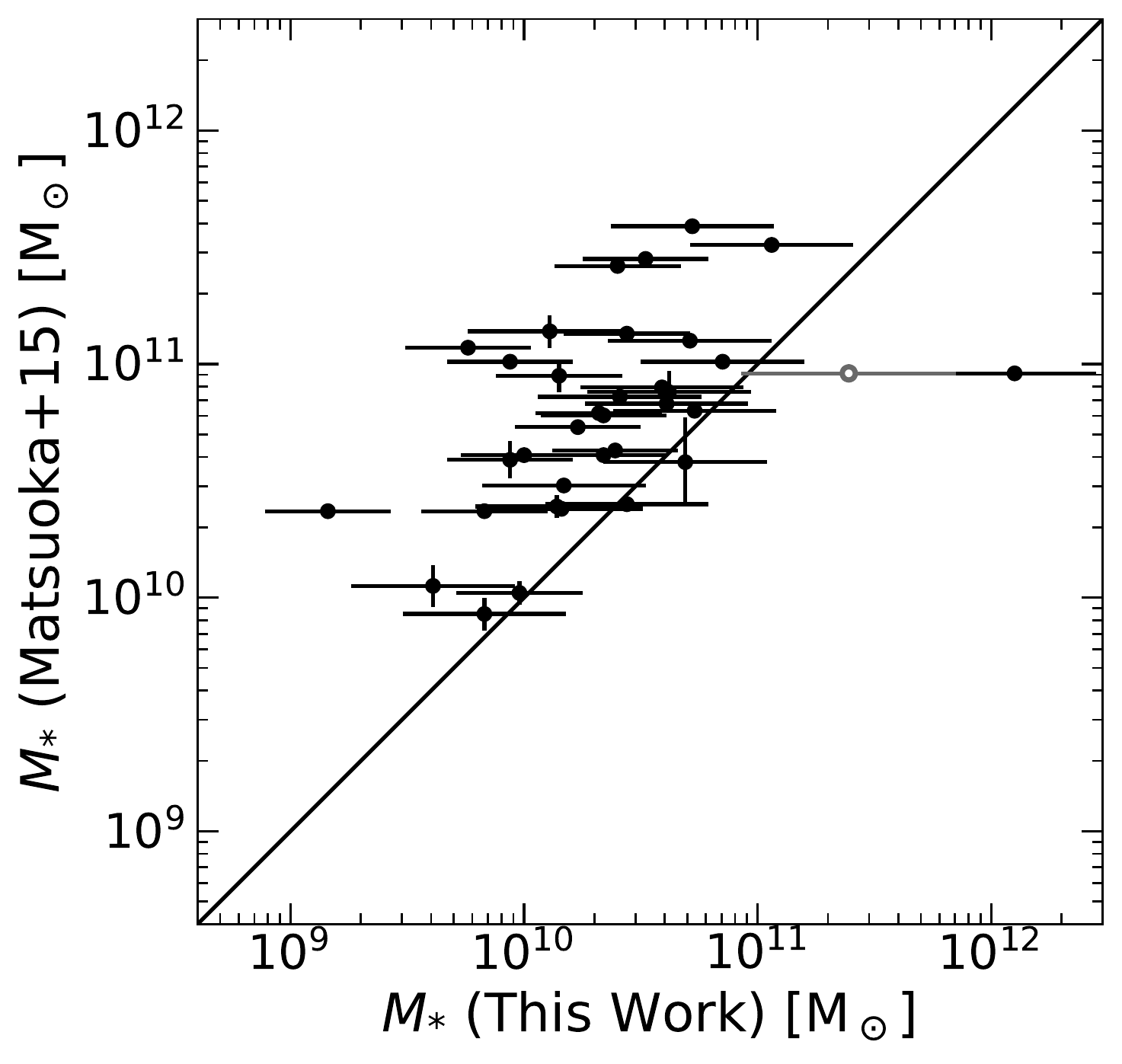}
\caption{Left: Comparison of the derived host light fraction from this work and \cite{Shen_etal_2015_Msigma}, both measured in the UVIS (F606W or F814W) bandpass. The black dashed line shows the 1:1 ratio line. Right: Comparison of the total stellar masses derived from this work and \cite{Matsuoka_etal_2015} for overlapping objects.}
\label{fig:fhost}
\end{figure*}

\subsection{$M_{\rm BH}-M_{\rm *, host}$ and $M_{\rm BH}-M_{\rm *, bulge}$ Relations of our sample}

Our sample consists of 38 sources spanning more than two orders of magnitude in $M_{\rm BH}$ and $M_{\rm *, host/bulge}$, which is sufficient for statistical analysis. {The Pearson correlation coefficient $r$ between $M_{\rm BH}$ and $M_{\rm *, host/bulge}$ are $r$\,$\sim$\,0.5 with low p-values ($<$0.05), suggesting the BH and galaxy/bulge masses are positively correlated and the correlation is statistically significant.} We use the {\tt LINMIX\_ERR} algorithm \citep{Kelly2007} to perform linear regression fitting on the $M_{\rm BH}-M_{\rm *,bulge}$ and $M_{\rm BH}-M_{\rm *, host}$ relations. {\tt LINMIX\_ERR} is a Bayesian fitting algorithm that accounts for uncertainties in both axes and intrinsic scatter in the relations. {We fit for the equation:
\begin{equation}\label{eq:true}
    {\rm log}\frac{M_{\rm BH}}{M_{\rm \odot}}=a+b\times {\rm log}\bigg({\frac{M_{\rm *}}{10^{10}M_{\rm \odot}}}\bigg)\ ,
\end{equation}}
and tabulate the best-fit parameters in Table \ref{tab:bestfit}.

\begin{table}[t]
\centering
\caption{{Best-fit Parameters of the Scaling Relations}} 
\begin{tabular}{l|ccc}
\hline\hline
Scaling Relations                & a & b & $\sigma$ \\
\hline
Bulge (Original)   & $7.44^{+0.13}_{-0.16}$  & $1.18^{+0.76}_{-0.52}$  & $0.39^{+0.11}_{-0.11}$      \\
Host  (Original)   & $7.34^{+0.14}_{-0.17}$  & $1.36^{+0.59}_{-0.40}$   & $0.34^{+0.11}_{-0.11}$      \\
\hline
Bulge (Bias-Corrected)    & $7.03^{+0.26}_{-0.41}$  & $1.67^{+0.83}_{-0.72}$  &  $0.59^{+0.23}_{-0.21}$      \\
Host (Bias-Corrected)      &  $7.01^{+0.23}_{-0.33}$  &  $1.74^{+0.64}_{-0.64}$ &  $0.47^{+0.24}_{-0.17}$     \\
\hline\hline
\end{tabular}
\label{tab:bestfit}
\end{table}

{The regression fits to the observed sample do not account for selection effects. In \S\ref{sec:mcmc} we use a more robust fitting code to constrain the intrinsic BH-host scaling relations, {and we include the bias-corrected best-fit parameters in Table \ref{tab:bestfit}.}. However, given the large dynamic range of our sample, selection effects do not appear to impact the results significantly, as we will show in \S\ref{sec:mcmc}. }

\subsection{Comparison with spectral decomposition}

We compare the host-light fraction and total stellar mass derived from HST imaging decomposition in this work and the spectral decompositions in \cite{Shen_etal_2015_Msigma} and \cite{Matsuoka_etal_2015}. Both of these earlier studies measured host-galaxy properties using the high-S/N coadded spectra from the first-year SDSS-RM spectra \citep{Shen_etal_2015_techoverview}. \cite{Shen_etal_2015_Msigma} used a principal component analysis method to decompose coadded spectra into quasar and galaxy spectra, and measured host-galaxy properties directly from the galaxy spectra, including stellar velocity dispersion and host-free AGN luminosity. \cite{Matsuoka_etal_2015} performed spectral decomposition on the coadded spectra using models of AGN and galaxy spectra, and measured host galaxy properties by fitting the decomposed galaxy spectra with stellar population models.

To compare the host-light fraction ($f_{\rm *}$, the fractional contribution of the host stellar component to the total flux), we calculate the host fraction in the HST imaging using the decomposed {\tt GALFIT} models within the 2$''$ diameter
spectral aperture, and the host fraction in spectral decomposition by computing the expected flux in the F606W and F814W bandpass in the decomposed spectra from \cite{Shen_etal_2015_Msigma}. Figure \ref{fig:fhost} (left panel) reveals that the host fraction from image decomposition is systematically higher than that derived from spectral decomposition, similar to our finding in the pilot study \citep{Li_etal_2021} and in \cite{Yue_etal_2018}. Figure \ref{fig:fhost} (right panel) compares the host stellar mass derived from this work and from \cite{Matsuoka_etal_2015}. Our stellar mass is systematically smaller by $\sim$0.5\,dex. {The cause of the stellar mass offset is currently unclear, but it might be partially due to different choices of initial mass functions and stellar population models in \cite{Matsuoka_etal_2015} and {\tt CIGALE}, or the fiber-loss correction applied in \cite{Matsuoka_etal_2015}, which assumes the mass-to-luminosity ratio in the central region (within the 2$''$-diameter aperture) represents that for the entire galaxy.}

\begin{figure*}[t]
\centering
\includegraphics[width=0.48\textwidth]{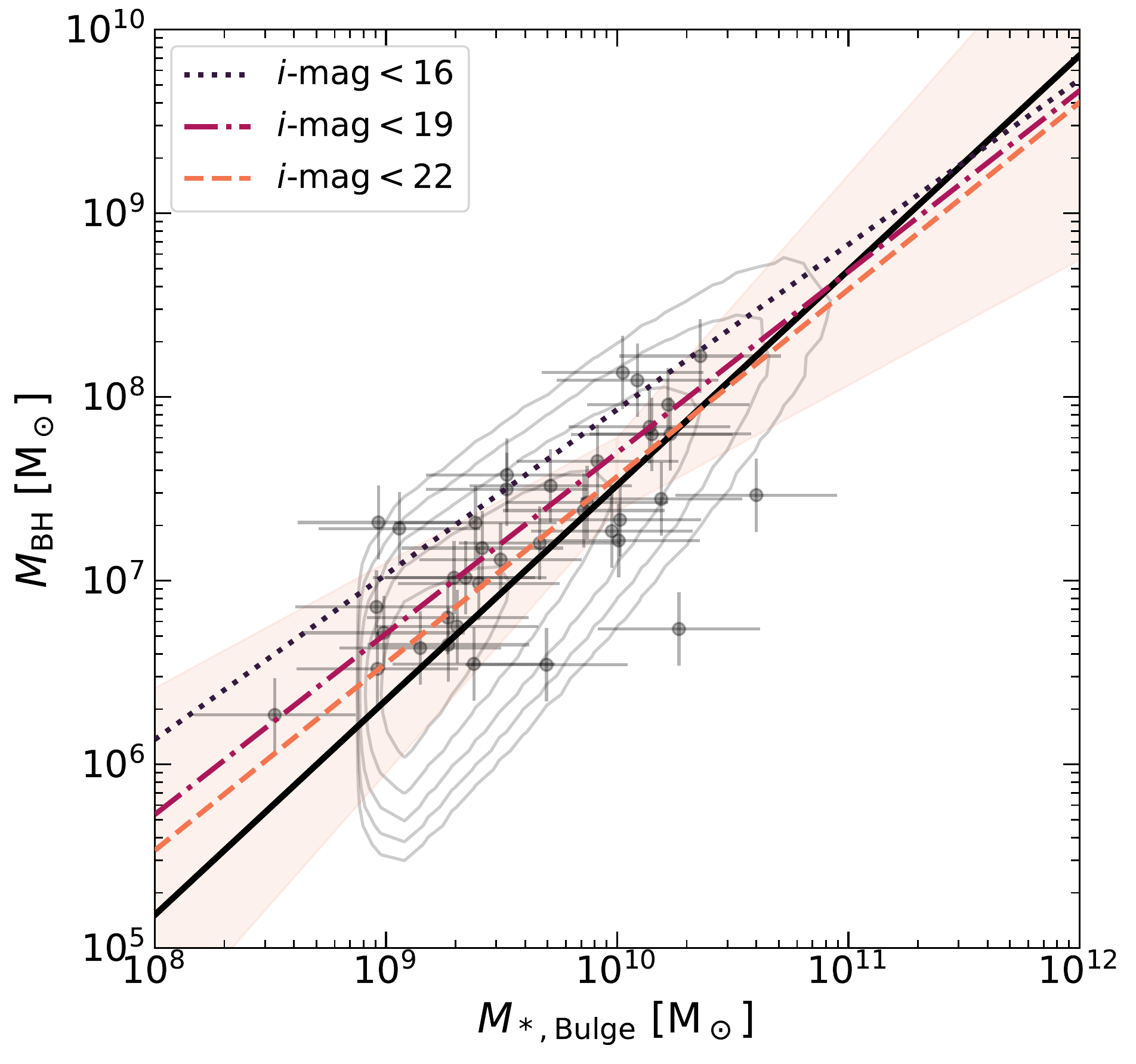}
\includegraphics[width=0.45\textwidth]{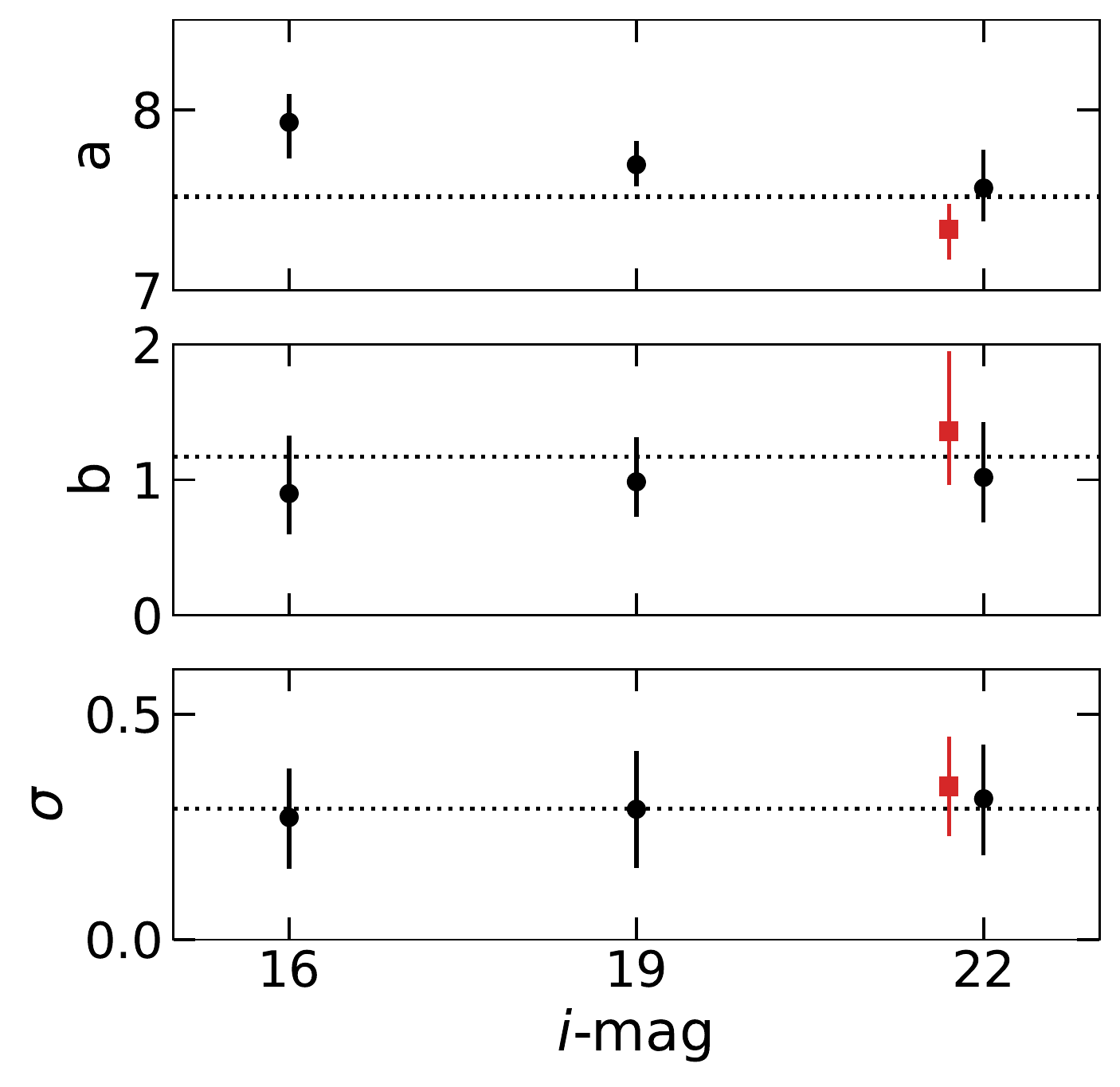}
\caption{Results of the \cite{Lauer_etal_2007} bias simulation described in Section \ref{sec:bias}. Left: grey contours represent the underlying quasar sample, and the grey points are one realization of the mock sample at the $i=22$ flux limit, after being perturbed with measurement uncertainties. The black solid line is the unbiased \cite{Kormendy_Ho_2013} relation. The {colored} lines indicate the results of the mock sample at different flux limits. Right: the best-fit $M_{\rm BH}-M_{\rm *, bulge}$ parameters at each flux threshold and the best-fit $M_{\rm BH}-M_{\rm *, host}$ parameters of our sample (plotted in red at $i=21.7$). The dotted lines show the parameters and intrinsic scatter from the \cite{Kormendy_Ho_2013} relation.}
\label{fig:bias}
\end{figure*}
\section{Discussion}\label{sec:discussion}

\subsection{Biases in the Observed BH Scaling Relations}\label{sec:bias}

{Since our sample is based on RM BH masses, it avoids the large statistical biases associated with the systematic uncertainties of SE masses \citep{Shen_Kelly_2010}.} To illustrate selection biases in our flux-limited sample due to intrinsic scatter in BH-host scaling relations \citep{Lauer_etal_2007}, we perform a forward-modeling simulation following the procedures in \cite{Shen_etal_2015_Msigma}. We first simulate a parent quasar sample following the local {\rm $M_{\rm *}$} distribution from \cite{Bernardi_etal_2010} and the {\rm $M_{\rm BH}-M_{\rm *, bulge}$} relation from \cite{Kormendy_Ho_2013}, with an intrinsic scatter of 0.29\,dex. Using the true {\rm $M_{\rm BH}$}, we assign a quasar bolometric luminosity by assuming a lognormal Eddington ratio distribution ($\lambda \equiv L_{\rm bol}/L_{\rm Edd}$) and the Eddington luminosity is $L_{\rm Edd} =1.26\times10^{38}(M_{\rm BH}/M_{\odot})\,{\rm erg\,s^{-1}}$. We choose a mean Eddington ratio of $\langle {\rm log} \lambda \rangle = -1$ and a scatter of 0.3\,dex \citep{Shen_etal_2008, Shen_Kelly_2012}. We include measurement uncertainties of 0.35\,dex for {\rm $M_{\rm *}$} and 0.2\,dex for {\rm $M_{\rm BH}$} to mimic the uncertainty levels of our CMLR-based {\rm $M_{\rm *}$} and RM {\rm $M_{\rm BH}$} measurements. Finally, for 100 bootstrap iterations, {we randomly draw 38 sources to perform {\tt LINMIX\_ERR} fitting with at different $i$-mag\,$<$ 16, 19, and 22 (similar to the flux limit of the SDSS-RM sample).

Figure \ref{fig:bias} shows how the flux limit biases the observed scaling relations. When the flux limit increases, over-massive BHs are preferentially selected, the slope of the best-fit relation becomes shallower, and the normalization increases}. The best-fit intrinsic scatter remains roughly the same in our simulations. However, our simulation does not include the outlier population with under-massive BHs seen in observations {(e.g., see Figure \ref{fig:Ms_Mgal_lit})}. Missing the outlier population could lead to an underestimation of the intrinsic scatter {and selection bias based on the flux limit}, since under-massive BHs are less likely to be selected in flux-limited surveys.  Given the relatively faint flux limit of our SDSS-RM sample, selection biases do not play an important role in the measured $M_{\rm BH}-M_{*}$ relations {(see the right panel of Figure \ref{fig:bias})}, and we found similar relations at $z_{\rm med}=0.5$ as the local relations. Our results are consistent with other studies that properly account for selection biases \citep[e.g.,][]{ Sexton_etal_2019, Suh_etal_2020, LiJunyao_etal_2021b}. 
{We note that the selection of our sample also depends on successful RM lag measurements, which may depend on BH mass and Eddington ratio, etc. However, since the lag-detection fraction in the \cite{Grier_etal_2017} sample is nearly uniform up to $z\sim$\,0.8, we assume the sample selection is not strongly affected by additional selection biases based on the quasar properties.}

\begin{figure*}[t]
\centering
\includegraphics[width=0.50\textwidth]{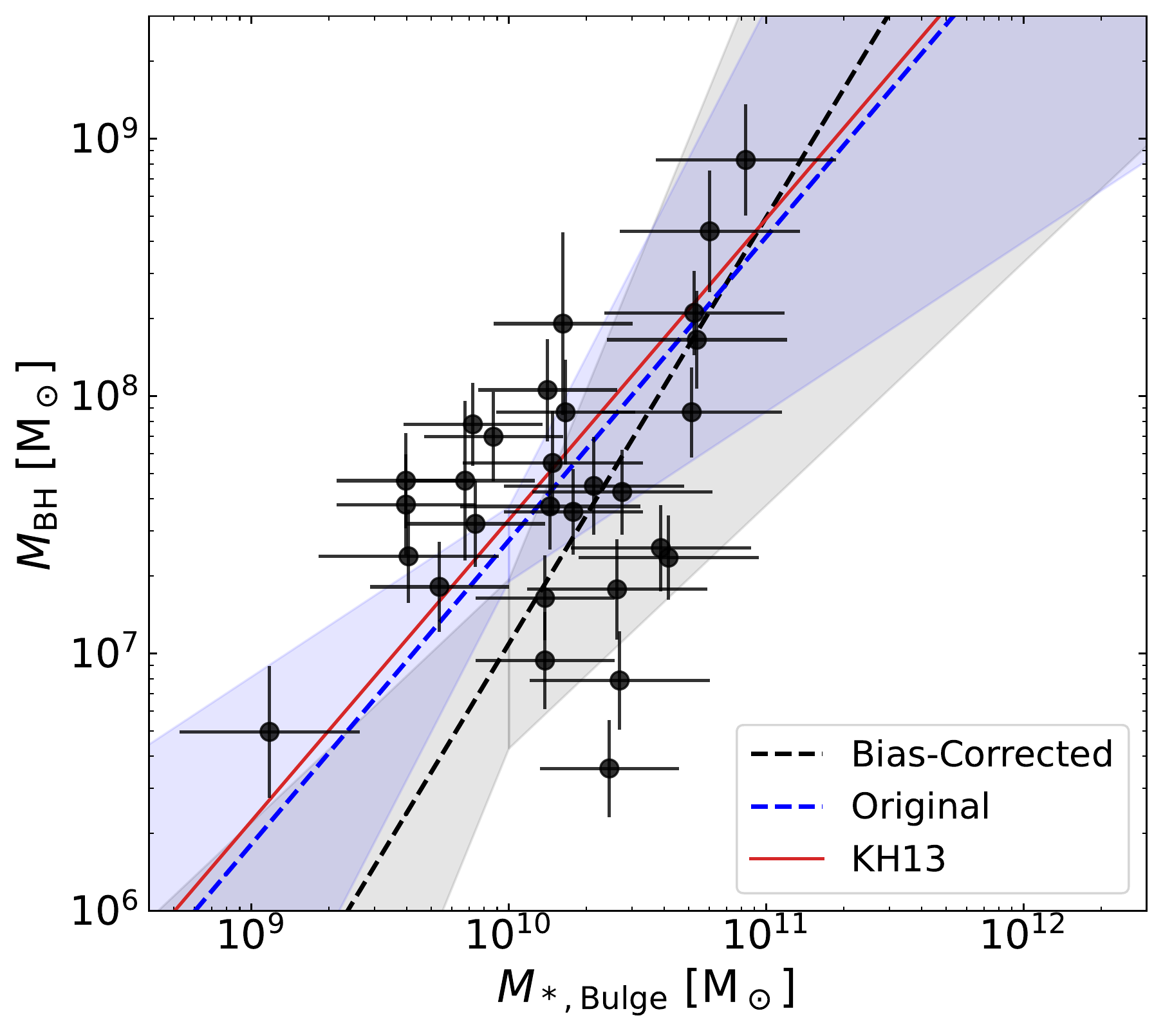}
\includegraphics[width=0.46\textwidth]{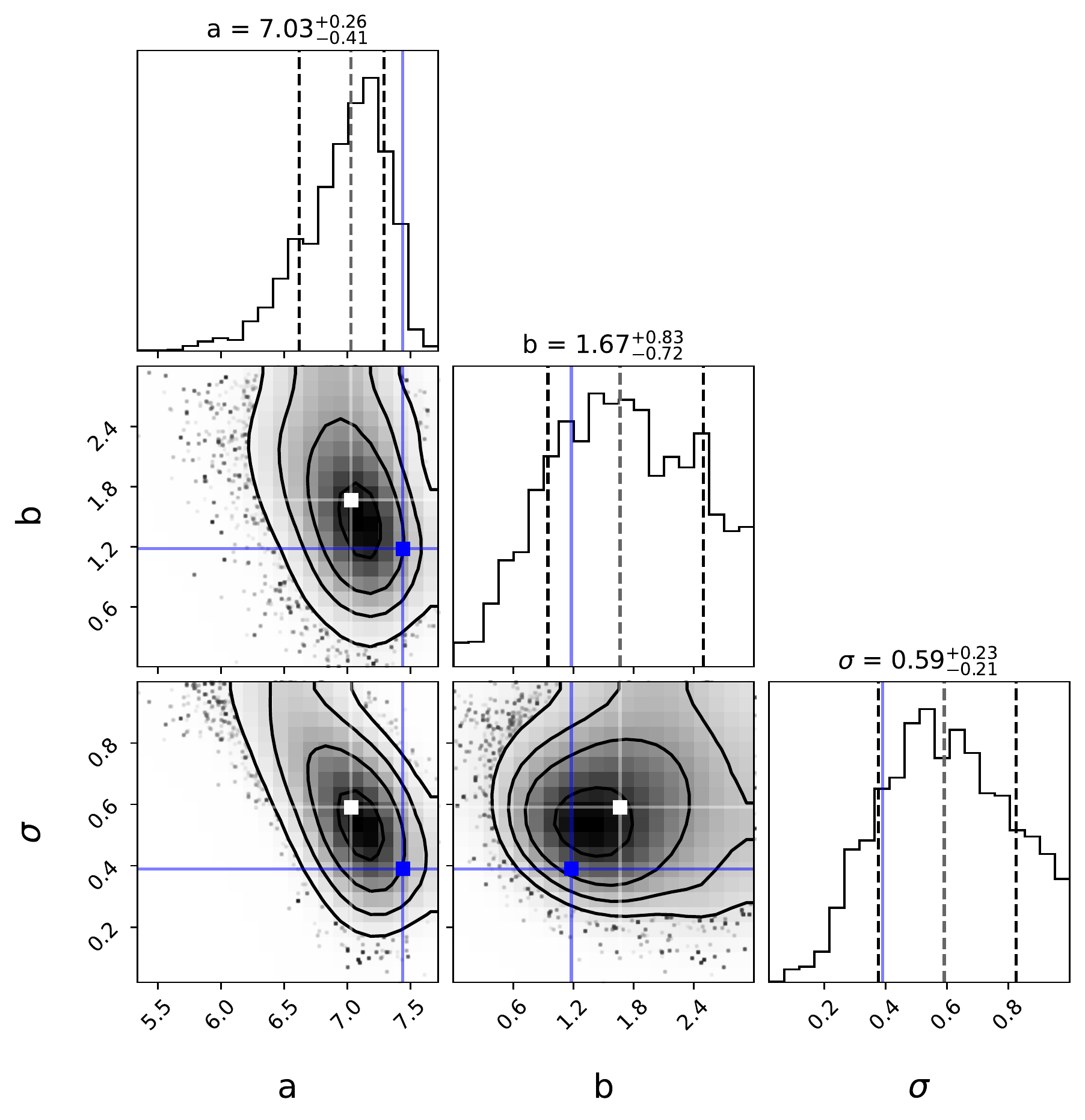}
\caption{MCMC fitting for the $M_{\rm BH}-M_{\rm *, bulge}$ relation when considering the selection bias. Left: the $M_{\rm BH}-M_{\rm *, bulge}$ relation with the 1$\sigma$ range drawn from the posterior (black dashed line and shaded area). The blue dashed lines and shaded area show the original best-fit relations from {\tt LINMIX\_ERR} (previously shown in Figure \ref{fig:Ms_Mgal}), and the red solid line is the local $M_{\rm BH}-M_{\rm *, bulge}$ from \cite{Kormendy_Ho_2013}. Right: the posterior distribution of the parameters $a$, $b$, and $\sigma$. {The best-fit parameters from {\tt LINMIX\_ERR} and the MCMC fitting are indicated by the blue and white squares, respectively. The MCMC fitting recovers a similar BH-galaxy relation to the original linear regression, indicating that the selection effects are minimal.}}\label{fig:mcmc_bulge}
\end{figure*}

\begin{figure*}[t]
\centering
\includegraphics[width=0.50\textwidth]{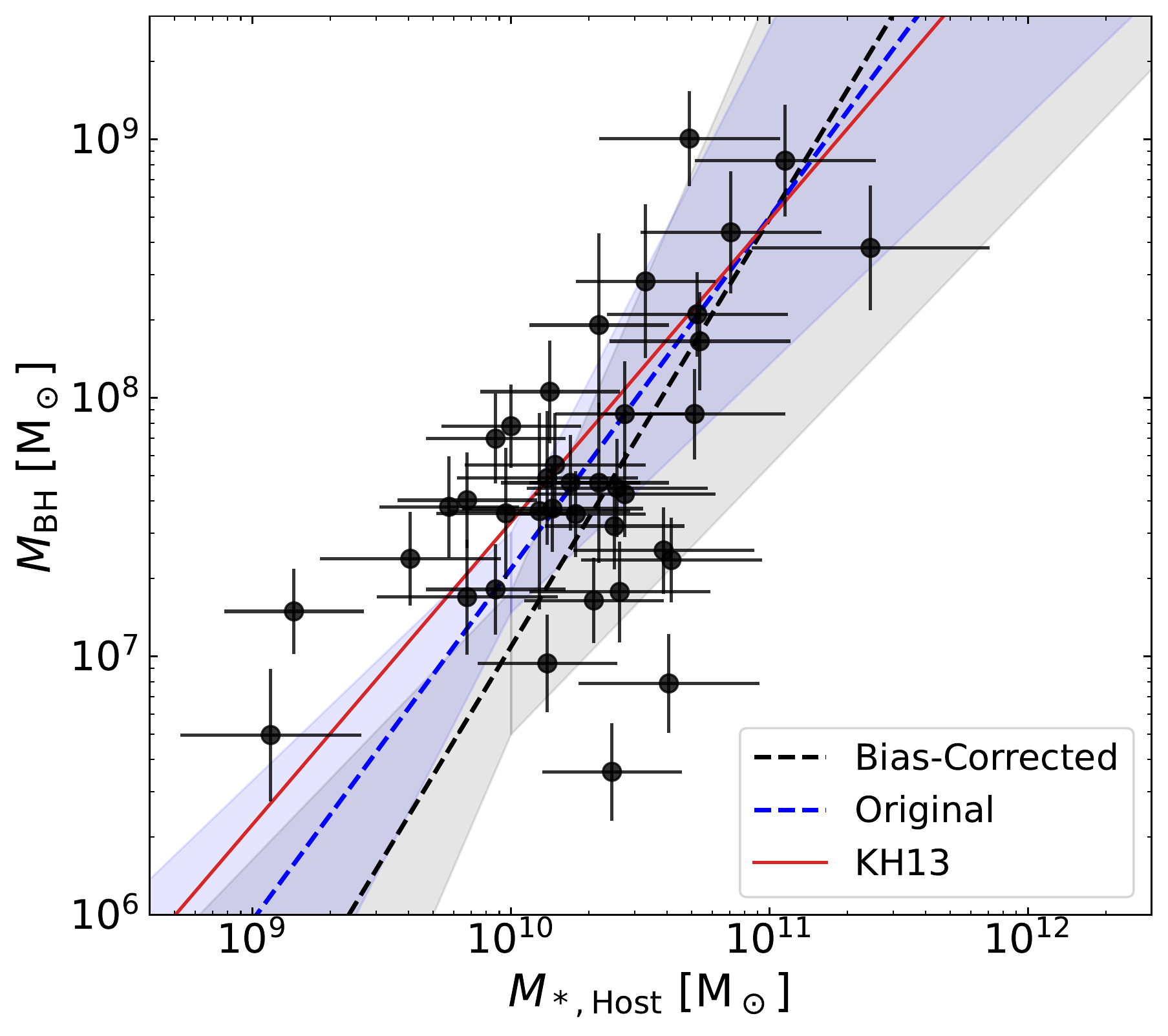}
\includegraphics[width=0.46\textwidth]{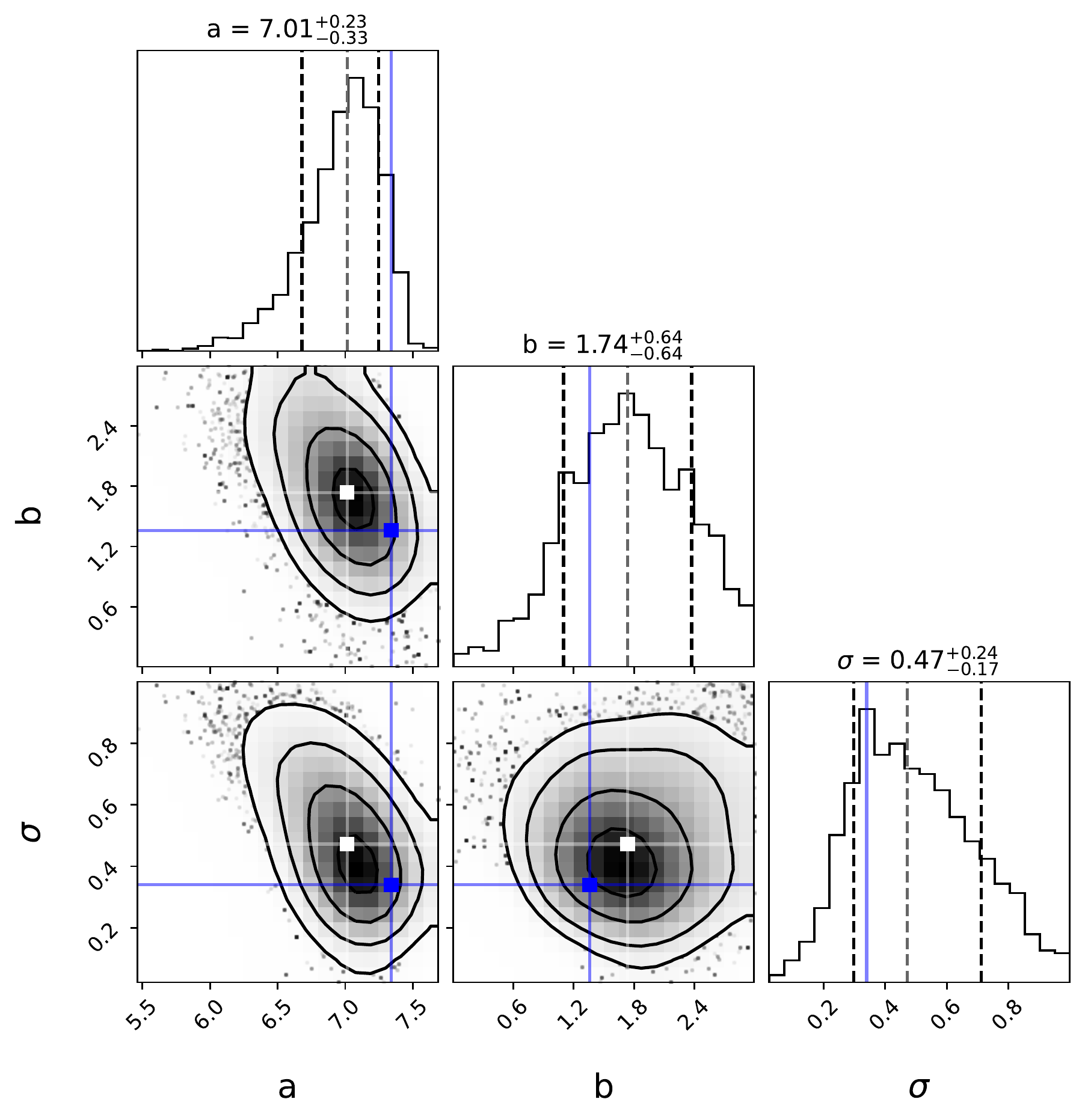}
\caption{Same format as Figure \ref{fig:mcmc_bulge} but for the $M_{\rm BH}-M_{\rm *, host}$ relation. } 
\label{fig:mcmc_host}
\end{figure*}

\subsection{Quantifying the Selection Effects}\label{sec:mcmc}

To quantitatively account for the underlying galaxy properties (i.e., the galaxy mass function) and selection effects, we follow the framework of \cite{Kelly2007} to perform a Markov chain Monte Carlo (MCMC) fitting for the intrinsic scaling relations and scatters. Specifically, we wrote a MCMC fitting code based on the Metropolis-Hastings algorithm \citep{Metropolis_etal_1953, Hastings_1970} and the statistical derivation in Section 4.1 and 5.1 of \cite{Kelly2007}, which allows us to account for the selection effect based on the dependent variable. {Here, we briefly summarize the fitting algorithm and parameter setup and refer the readers to \cite{Kelly2007} for the full, detailed mathematical derivation. Our custom fitting code is available via \url{ftp://quasar.astro.illinois.edu/public/sdssrm/paper_data/Li_2023_HST_host}.}

As shown in \cite{Kelly2007}, when the intrinsic scatter and the uncertainties are comparable to the dynamical range of the data, the best-fit slope becomes shallower when the underlying distribution is not considered as a prior in the fitting procedure. One solution is to incorporate empirical distributions into the likelihood function, e.g., the observed local stellar mass function $\Phi(x)$ from literature \citep[e.g.,][]{Bernardi_etal_2010}. However, due to the limited dynamical range in $M_{\rm *}$ and the small sample size of our data, the local galaxy mass function is not a good prior for our sample. Alternatively, \cite{Kelly2007} suggests using a series of Gaussian functions to model the underlying distribution, which provides a flexible and empirical solution even when the underlying distribution is unknown. This is the method implemented in the original {\tt LINMIX\_ERR} fitting algorithm, which we continue to adopt in our MCMC fitting for consistency.

When the sample selection is based on the dependent variable (i.e., $M_{\rm BH}$), the posterior distribution and likelihood depend on an additional term ($P(I=1|\theta)$, where $\theta$ are the model parameters) that describes the likelihood of including each data point in the observed sample based on the model parameters \citep[for more details, see Section 5.1 in][]{Kelly2007}. 
Following the same procedure as described in Paragraph 1 of this section, we estimate the expected $i$-band magnitude by assigning a random Eddington ratio and an Eddington luminosity based on the redshift and a range of ``true'' $M_{\rm BH}$ for each data point. The probability of including a data point is 1 if $i$-mag$<21.7$ and 0 if $i$-mag$>21.7$. Finally, we calculate $P(I=1|\theta)$ by integrating the probability of including each data point and their likelihood over a range of ``true'' $M_{\rm BH}$ and $M_{\rm *}$ given the model parameters. We adopt uninformative, flat priors for all parameters ($5<a<10$, $0<b<3$, and $0.0001<\sigma^{2}<1$) and minimize the product of the likelihood and prior to compute the posterior distribution of the parameters $a$, $b$, and $\sigma$.

Figure \ref{fig:mcmc_bulge} and \ref{fig:mcmc_host} present the posterior distribution of  $a$, $b$, and $\sigma$, and the best-fit values are tabulated in Table \ref{tab:bestfit}. As expected, the slope of the intrinsic scaling relations becomes steeper, and the normalization decreases, after correcting for the selection biases. The best-fit parameters are within uncertainties as the {\tt LINMIX\_ERR} fit (without considering selection bias) discussed in Section \ref{sec:results}, demonstrating that our results are not strongly affected by selection biases.

Intrinsic scatter of the $M_{\rm BH}-M_{\rm *,bulge}$ and $M_{\rm BH}-M_{\rm *, host}$ relations is an important indicator for BH$-$galaxy co-evolution, as it might be related to the galaxy/AGN properties and their evolutionary path. The local samples of \cite{Kormendy_Ho_2013} and \cite{Bennert_etal_2021} only include classical bulges and pseudo-bulges and have a smaller intrinsic scatter of $0.28-0.39$\,dex. However, when including all morphological types and active/inactive galaxies, the intrinsic scatter increases to $\sim$0.5\,dex \citep{Reines_Volonterni_2015, Bentz_etal_2018} for the $M_{\rm BH}-M_{\rm *, bulge}$ relation, and becomes even slightly larger for the $M_{\rm BH}-M_{\rm *, host}$ relation.
{In addition, the BH accretion rate is found to be correlated with other host properties, e.g., compactness of the central $\sim$1\,kpc region \citep{Ni_etal_2019,Ni_etal_2021}, which can introduce additional scatter in the BH scaling relations.}

For our quasar sample, the intrinsic scatter of the $M_{\rm BH}-M_{\rm *, host}$ and $M_{\rm BH}-M_{\rm *, bulge}$ relations are {$0.47_{-0.17}^{+0.24}$\,dex and $0.59_{-0.21}^{+0.23}$\,dex}, respectively, after accounting for the selection effects, which are comparable to the scatter in the local relations. The intrinsic scatter of our $M_{\rm BH}-M_{\rm *, host}$ relation is slightly smaller ($\lesssim\,0.5\sigma$ of difference) than our $M_{\rm BH}-M_{\rm *, bulge}$ relation, which {we will further discuss in Section \ref{sec:bulge_vs_host}}. {Because we have neglected the systematic uncertainty in our RM BH masses due to the scatter in individual virial coefficients, the actual intrinsic scatter in the BH-host stellar mass relations for $0.2<z<0.8$ quasars might be even smaller. }

\begin{figure*}
\centering
\includegraphics[width=0.8\textwidth]{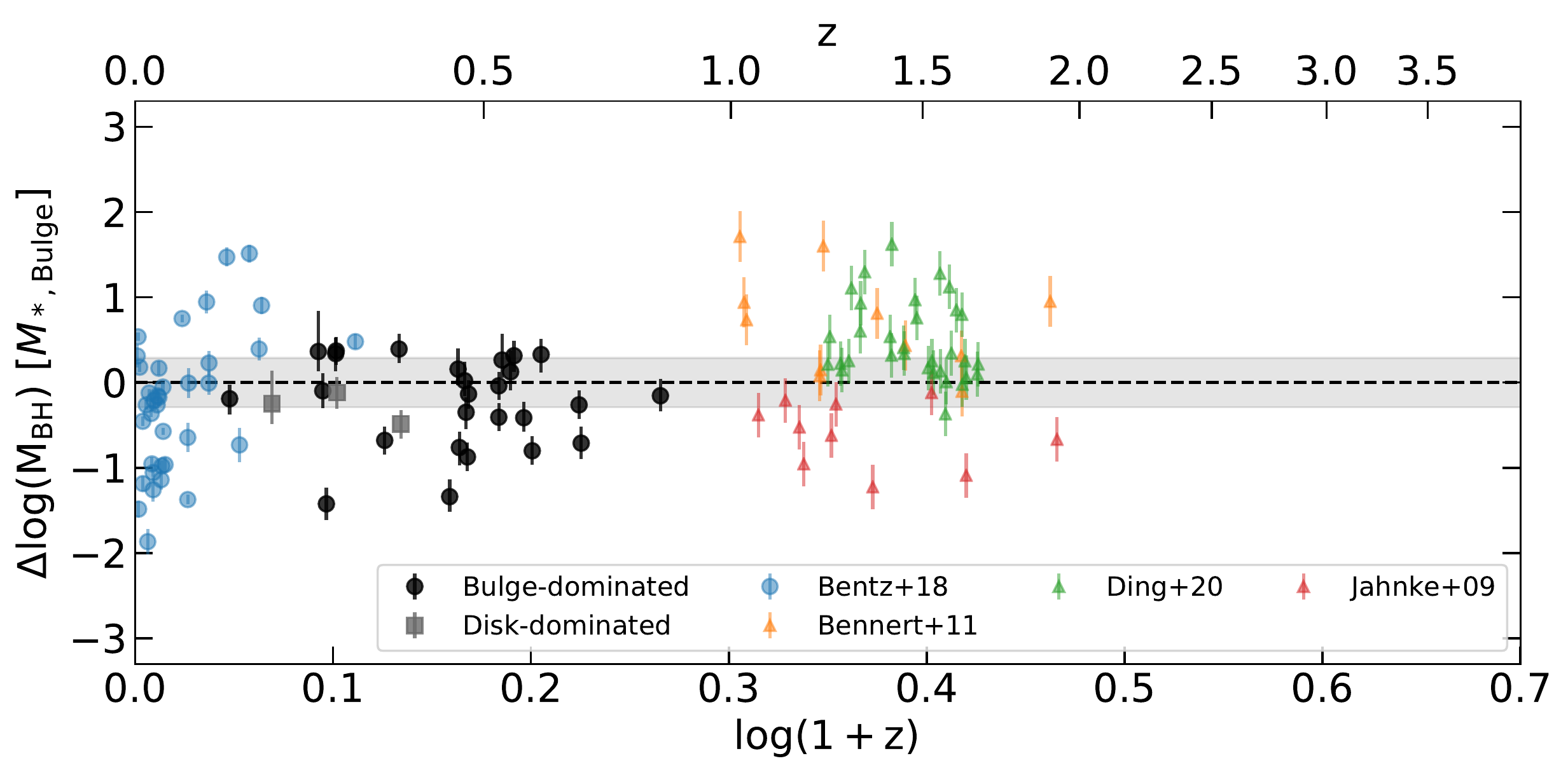}\\
\includegraphics[width=0.8\textwidth]{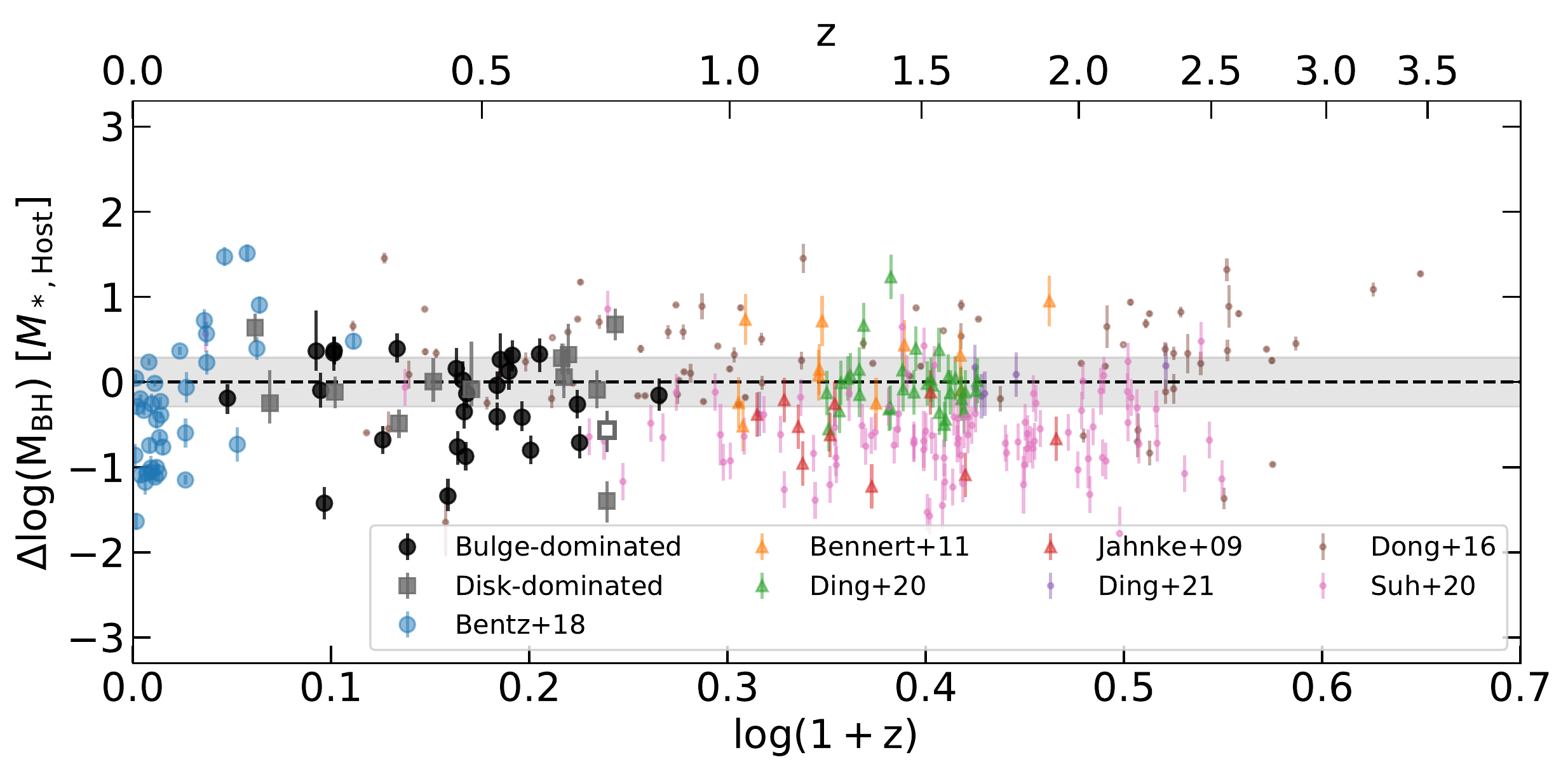}
\caption{{Evolution of ${\rm \Delta log(M_{BH})}$ with redshift, with baselines adopted from the best-fit relations of $M_{\rm BH}-M_{\rm Bulge}$ and $M_{\rm BH}-L_{\rm Bulge}$ from the \cite{Kormendy_Ho_2013} sample. {Our work is the only sample with RM-based BH masses beyond $z>0.3$.} Vertical error bars are from uncertainties in BH mass only.} }
\label{fig:redshift}
\end{figure*}

\subsection{Evolution of BH Scaling Relations}

Earlier works on $M_{\rm BH}-M_{\rm *, bulge}$ and $M_{\rm BH}-M_{\rm *, host}$ relations found that the average BH-to-host galaxy mass ratio evolves positively with redshift \citep{Peng_etal_2006a, Peng_etal_2006b, Merloni_etal_2010, Bernardi_etal_2010}. However, selection biases and measurement uncertainties could yield false positives of the evolution. For example, \cite{Jahnke_etal_2009} reported that there is no evidence of evolution when they carefully choose their sample to avoid selection biases. Similarly, \cite{Suh_etal_2020} found the redshift evolution seen in \cite{Merloni_etal_2010} can be explained by {the Lauer bias \citep{Lauer_etal_2007}, and there is no trend of evolution in their X-ray-selected, lower luminosity sample}. Using over 500 uniformly-selected $0.2<z<0.8$ SDSS quasars, \cite{LiJunyao_etal_2021b} found a redshift evolution of the offset in the $M_{\rm BH}-M_{\rm *, host}$ relation that is within $\pm$0.2\,dex from zero, consistent with no significant evolution since $z\sim0.8$.

Figure \ref{fig:redshift} presents the deviation of $M_{\rm BH}$ from the \cite{Kormendy_Ho_2013} {$M_{\rm BH}-M_{\rm *, bulge}$} relation of our sample. We fit the deviation as a function of ${\rm log({\it z})}$, and the slopes and intercepts are consistent with zero, suggesting there is no redshift evolution from the local relations. The {median (16th/84th percentiles)} black hole/host galaxy mass ratios are {${M_{\rm BH}}/{M_{\rm *, bulge}}=0.0037 (0.0007/0.0088)$ and ${M_{\rm BH}}/{M_{\rm *, host}}=0.0030 (0.0007/0.0076)$}, {within 1$\sigma$ uncertainty} of the local value of {${M_{\rm BH}}/{M_{\rm *, bulge}}\sim0.005$} \citep{Kormendy_Ho_2013}. 

Our results agree with recent observational studies that there is limited evolution in the $M_{\rm BH}-M_{\rm *}$ relation {\citep{Suh_etal_2020,LiJunyao_etal_2021b}}. {The limited redshift evolution in the $M_{\rm BH}-M_{\rm *}$ relations can be explained by the tight correlation between the BH accretion rates and star formation rates found in bulge-dominated galaxies at $z=0.5-3$ \citep{Yang_etal_2019}, which suggests the growth of SMBH and their host galaxies and bulges are in sync since $z\sim3$.}

\subsection{BH$-$Bulge versus BH$-$Host Relations}\label{sec:bulge_vs_host}
In the local Universe, SMBH masses are tightly correlated with the properties of classical bulges, but not with disks or total mass of the host galaxy. However, the intrinsic scatter in the $M_{\rm BH}-M_{\rm *, host}$ relation of our sample is slightly smaller than that in the $M_{\rm BH}-M_{\rm *,bulge}$ relation. There are a few extra sources of uncertainties for the bulge-mass estimate than for the total mass estimate, {which will contribute to the intrinsic scatter}. First of all, bulge-disk decomposition could add significant uncertainties to the bulge mass. Most of our host galaxies are far less luminous than the quasar, and compact hosts could be near the limit of HST imaging resolution. Bulge-disk decomposition is reliable when both the bulge and disk are sufficiently bright (compared to the central {quasar}) and there is a distinct difference in their effective radii. Furthermore, we cannot distinguish classical bulges from pseudo-bulges or other bulge-like structures from surface-brightness decomposition, nor could we model complex bar, spiral, and merger structures in the bulge-disk decomposition, which increases the uncertainties in bulge identification and mass estimation. {\cite{Gao_etal_2017} found that rigorous modeling of bars and innermost structures (e.g., rings and disk breaks near the bulge) is crucial to recovering bulge properties, while the modeling of spiral arms and extended disks have negligible effects. We note that some host galaxies in our sample show clear evidence of bars (e.g., RM320, RM634, etc), which are modeled as bulges ($n=4$) or disks ($n=1$) in our analysis, without additional bar structures. Moreover, some host galaxies show clear spiral arm features (e.g., RM371, RM772, etc), indicating the presence of disks, but are modeled as ``bulges''($n=4$). Previous works \cite[e.g.,][]{Zhao_etal_2021,Greene_etal_2020} showed late-type quasar hosts preferentially scatter below early-type hosts in the $M_{\rm BH}-M_{\rm *,bulge}$ relation, which is not seen in our data, suggesting our bulge-disk decomposition is not as reliable in measuring galaxy morphology.} A detailed simulation of bulge-disk decomposition for AGNs with similar host and quasar properties (e.g., AGN/host flux ratio, host effective radius, S\'ersic indices, and complex structures) is needed to provide quantitative uncertainty estimation, which is beyond the scope of this work. 

Another possible source of uncertainty is the CMLR estimation for bulges. Recent studies have reported that compact regions around the SMBH may have denser interstellar medium, boosted star formation, and complex stellar populations \citep[e.g.,][]{Ni_etal_2019, Kim_etal_2019, Zhuang_etal_2020, Shangguan_etal_2020, Yesuf_etal_2020, Molina_etal_2021}. Two-band color and the use of empirical M/L relation may not be sufficient to produce reliable estimates for the bulge stellar mass.

\section{Conclusions}\label{sec:conclusions}
We present the $M_{\rm BH}-M_{\rm *, bulge}$ and $M_{\rm BH}-M_{\rm *, host}$ relations of 38 sources with RM-based BH masses \citep[][]{Grier_etal_2017} and $0.2\lesssim z\lesssim 0.8$ (median redshift $z_{\rm med}=0.5$). Our sample is the first uniformly-selected sample with RM-based BH masses at $z>0.3$ for studying BH-host relations, and covers two orders of magnitude in BH mass and host stellar mass. The reliable RM-based BH masses and host mass estimates from HST imaging decomposition, combined with the large sample size and dynamic range in mass, allow one to alleviate selection biases in studying the potential evolution of the BH-host scaling relations. {Our scaling relations are consistent with those for local AGNs, quiescent galaxies, and other high-redshift samples, with negligible redshift evolution up to $z\lesssim 1$. As shown in Table \ref{tab:bestfit}, the best-fitting intrinsic $M_{\rm BH}-M_{\rm *,host}$ relation is: $\log (M_{\rm BH}/M_{\rm \odot})=7.01_{-0.33}^{+0.23} + 1.74_{-0.64}^{+0.64}\log (M_{\rm *,host}/10^{10}M_{\rm \odot})$ after correcting for the underlying sample distribution and selection effects. We estimate an intrinsic scatter of $0.59^{+0.23}_{-0.21}$\,dex and $0.47^{+0.24}_{-0.17}$\,dex in the $M_{\rm BH}-M_{\rm *, bulge}$ and $M_{\rm BH}-M_{\rm *, host}$ relations, respectively, which is again consistent with the local BH scaling relations. } {With our approved Cycle 1 JWST proposal (GO-2057, PI: Shen), we will continue to explore BH$-$host relations and their redshift evolution up to $z\sim2$ using quasars with direct RM-based BH masses \citep{Grier_etal_2019}.}

\begin{acknowledgments}
JIL acknowledges support from the Government Scholarship to Study Abroad (GSSA) from the Ministry of Education of Taiwan and support from the Illinois Space Grant Consortium (ISGC) Graduate Fellowship. 
YS acknowledges support from NSF grants AST-1715579 and  AST-2009947. 
LCH was supported by the National Science Foundation of China (11721303, 11991052, 12233001, 12011540375) and the China Manned Space Project (CMS-CSST-2021-A04, CMS-CSST-2021-A06).
WNB acknowledges support from NSF grant AST-2106990.
PBH is supported by NSERC grant 2017-05983.
Based on observations with the NASA/ESA Hubble Space Telescope obtained from the Data Archive at the Space Telescope Science Institute, which is operated by the Association of Universities for Research in Astronomy, Incorporated, under NASA contract NAS5-26555. Support for Program number HST-GO-15849 was provided through a grant from the STScI under NASA contract NAS5-26555. 
\end{acknowledgments}

\software{
AstroDrizzle, 
Astropy \citep{astropy:2013,astropy:2018}, 
CIGALE \citep{Boquien_etal_2019}, 
{\tt GALFIT} \citep{galfit}, 
{\tt LINMIX\_ERR} \citep{Kelly2007},
matplotlib \citep{matplotlib}, 
Numpy \citep{numpy}, 
{\tt photutils} \citep{photutils}, 
{\tt pysynphot} \citep{pysynphot}, 
seaborn \citep{seaborn}.}

\facilities{HST (WFC3/UVIS, WFC3/IR)}

\bibliography{refs}

\end{document}